\documentstyle[preprint,eqsecnum,pre,aps,epsf,epsfig]{revtex}

\begin{document}
\draft

\title{Diluted Networks of Nonlinear Resistors and Fractal Dimensions of Percolation Clusters
}
\author{H. K. Janssen and O. Stenull
}
\address{
Institut f\"{u}r Theoretische Physik 
III\\Heinrich-Heine-Universit\"{a}t\\Universit\"{a}tsstra{\ss}e 1\\40225 
D\"{u}sseldorf, Germany
}
\date{\today}
\maketitle

\tightenlines

\begin{abstract}
We study random networks of nonlinear resistors, which obey a generalized Ohm's law, $V\sim I^r$. Our renormalized field theory, 
which thrives on an interpretation of the involved Feynman Diagrams as being 
resistor networks themselves, is presented in detail. By considering distinct values of the nonlinearity $r$, we calculate several fractal dimensions characterizing percolation clusters. For the dimension associated with the red bonds we show that $d_{\mbox{{\scriptsize red}}} = 1/\nu$ at least to order ${\sl O} \left( \epsilon^4  \right)$, with $\nu$ being the correlation length 
exponent, and $\epsilon = 6-d$, where $d$ denotes the spatial dimension. This 
result agrees with a rigorous one by Coniglio. Our result for the chemical distance, $d_{\mbox{{\scriptsize 
min}}} = 2 - \epsilon /6 - \left[ 937/588 + 45/49 \left( \ln 2 -9/10 \ln 3 
\right)\right] \left( \epsilon /6 \right)^2 + {\sl O} \left( \epsilon^3  
\right)$ verifies a previous calculation by one of us. For the backbone dimension we find $D_B = 2 + \epsilon /21 - 172 \epsilon^2 /9261 + 2 \left( - 74639 + 22680 \zeta \left( 3 \right) \right)\epsilon^3 /4084101 + {\sl O} \left( \epsilon^4  \right)$, where $\zeta \left( 3 \right) = 1.202057...$, in agreement to second order in $\epsilon$ with a two-loop calculation by Harris and Lubensky.
\end{abstract}
\pacs{PACS numbers: 64.60.Ak, 64.60.Fr, 72.80.Ng, 05.70.Jk}

\section{Introduction}
\label{introduction}

Percolation is a leading paradigm for disorder (for a review see 
e.g.\cite{bunde_havlin_91,stauffer_aharony_92,hughes_95}). Though it represents the simplest model of a disordered system, it has many applications e.g.\ 
polymerization, porous and amorphous materials, thin films, spreading of 
epidemics etc. In particular the transport properties of percolation clusters 
have gained a vast amount of interest over the last decades. Random resistor 
networks (RRN) play a major role in the study of transport on percolation 
clusters for several reasons. For example, one can learn about the conductivity of disordered media, which might be important for technical applications. One can study diffusion on disordered substrates, since the diffusion constant $D$ and the conductivity $\Sigma$ of the system are related by the 
Einstein equation
\begin{eqnarray}
\Sigma = \frac{e^2 n}{K_B T} D \ ,
\end{eqnarray}
where $e$ and $n$ denote the charge and the density of the mobile particles. 
Nonlinear RRN, for which the voltage drop $V$ over an individual resistor depends on some power $r$ of the current $I$ flowing through it, can be exploited to derive the fractal dimension of various substructures of percolation clusters.

In this paper we present in detail our study of nonlinear RRN by renormalized field theory. A brief account of this work has been given previously 
in\cite{janssen_stenull_oerding_99}. It is based on an approach by 
Stephen\cite{stephen_78}, its refinements by Harris and 
Lubensky\cite{harris_lubensky_84} and its generalization to nonlinear resistors by Harris\cite{harris_87}. Our work thrives on the interpretation of the 
involved Feynman Diagrams as being resistor networks 
themselves\cite{stenull_janssen_oerding_99,janssen_stenull_oerding_99}. This 
interpretation leads to a substantial simplification of the field theoretic 
calculations, as we demonstrate by calculating the fractal dimensions of the 
chemical length and the backbone to two- and three-loop order respectively.

\section{The Model}
\label{model}
Consider a $d$-dimensional lattice, where bonds between nearest neighboring sites are randomly occupied with 
probability $p$ or empty with probability $1-p$. Each occupied bond has a finite non zero conductance $\sigma$ whereas unoccupied bonds have conductance zero. We suppose that the system is near the percolation threshold, i.e.\ $p$ is close to the critical concentration $p_c$ above which an infinite cluster exists. We are interested in the resistance $R_r ( x , x^\prime )$ between two lattice 
sites $x$ and $x^\prime$ averaged subject to the condition, that $x$ and 
$x^\prime$ are on the same cluster,
\begin{eqnarray}
M_r = \langle \chi (x ,x^\prime) R_r (x ,x^\prime ) \rangle_C / \langle \chi (x 
,x^\prime) \rangle_C \ . 
\end{eqnarray}
$\langle ...\rangle_C$ denotes the average over all configurations of the 
diluted lattice and $\chi (x ,x^\prime)$ is an indicator function that takes the 
value one if $x$ and $x^\prime$ are on the same cluster and zero otherwise. Note that $\langle \chi (x ,x^\prime) \rangle_C$ is nothing more than the usual correlation function in percolation theory. At 
criticality $M_r$ obeys\cite{harris_fisch_77,dasgupta_harris_lubensky_78}
\begin{eqnarray}
\label{MbehavesAs}
M_r \sim | x - x^\prime |^{\phi_r / \nu} \ , 
\end{eqnarray}
where $\nu $ is the correlation length exponent defined by $\xi \sim 
(p-p_c)^{-\nu}$.

\subsection{Kirchhoff's laws}
\label{kirchhoffsLaws}
Here, we consider a nonlinear RRN as proposed by Kenkel and Straley\cite{kenkel_straley_82}. The bonds between nearest neighboring sites $i$ and $j$ obey a generalized Ohm's Law,
\begin{eqnarray}
V_j - V_i = \rho_{i,j} I_{i,j} |I_{i,j}|^{r-1} \ ,
\end{eqnarray}
or equivalently, 
\begin{eqnarray}
\sigma_{i,j} \left( V_j - V_i \right) | V_j - V_i |^{s-1} = I_{i,j} \ , 
\end{eqnarray}
where $\sigma_{i,j}$ ($\rho_{i,j}$) is the nonlinear conductance (resistance) of the bond $\left\langle i,j \right\rangle$, $I_{i,j}$ is the current flowing through the bond from $j$ to $i$ and $V_i$ is the 
potential at site $i$. The exponents $r$ and $s$ are describing the non linearity with $r=s^{-1}$. The conductance and the resistance are related via $\sigma_{i,j} = \rho_{i,j}^{-s}$.

Suppose a current $I$ is put into a cluster at site $x$ and taken out at site 
$x^\prime$. Those sites connected to $x$ and $x^\prime$ by mutually avoiding paths are constituting the backbone between $x$ and $x^\prime$. The 
power dissipated on the backbone is by definition
\begin{eqnarray}
\label{powerDef}
P=I \left( V_x - V_{x^\prime} \right) \ .
\end{eqnarray}
Using Ohm's law, it may be expressed entirely in terms of voltages as
\begin{eqnarray}
\label{powerInTermsOfV}
P=  R_r (x ,x^\prime)^{-1} \left| V_x - V_{x^\prime} \right|^{s+1} = \sum_{\langle i,j \rangle} 
\sigma_{i,j} |V_i - V_j|^{s+1} =  P \left( \left\{ V \right\} \right) \ ,
\end{eqnarray}
where the sum is taken over all nearest neighbor pairs on the cluster and 
$\left\{ V \right\}$ denotes the corresponding set of voltages. As a consequence of the variation principle
\begin{eqnarray}
\label{variationPrinciple1}
\frac{\partial}{\partial V_i} \left[ \frac{1}{s+1} P \left( \left\{ V \right\} 
\right) - \sum_j I_j V_j \right] = 0 \ ,
\end{eqnarray}
one obtains the circuit equations
\begin{eqnarray}
\label{cirquitEquations}
\sum_{\langle j \rangle} \sigma_{i,j} \left( V_i - V_j \right) |V_i - V_j |^{s-1}  = - \sum_{\langle j \rangle} I_{i,j} =I_i \ ,
\end{eqnarray}
where $I_i = I \left( \delta_{i,x} - \delta_{i,x^\prime} \right)$ and the summations extend over the nearest neighbors of $i$.

Alternatively to Eq.~(\ref{powerInTermsOfV}) the power can by rewritten in terms of the currents as
\begin{eqnarray}
\label{powerInTermsOfI}
P=  R_r (x ,x^\prime) |I|^{r+1} = \sum_{\langle i,j \rangle} \rho_{i,j} |I_{i,j}|^{r+1} =  P 
\left( \left\{ I \right\} \right) \ ,
\end{eqnarray}
with $\left\{ I \right\}$ denoting the set of currents flowing through the 
individual bonds. Obviously the cluster may contain closed loops as subnetworks. 
Suppose there are currents $\left\{ I^{(l)} \right\}$ circulating independently 
around a complete set of independent closed loops. Then the power is not only a function of $I$ but also 
of the set of loop currents. The potential drop around closed loops is zero. This gives rise to the variation principle
\begin{eqnarray}
\label{variationPrinciple2}
\frac{\partial}{\partial I^{(l)}} P \left( \left\{ I^{(l)} \right\} , I \right) 
= 0 \ .
\end{eqnarray} 
Eq.~(\ref{variationPrinciple2}) may be used to eliminate the loop currents and 
thus provides us with a method to determine the total resistance of the backbone 
via Eq.~(\ref{powerInTermsOfI}).

\subsection{Connection to cluster properties}
\label{clusterProperties}
Here we provide background on the meaning of $\phi_r$ for some specific values 
of $r$. For $r \to 1$, one recovers the linear RRN. $\phi_1$ is the usual 
resistance exponent as studied to order $\epsilon^2$, e.g., in\cite{stenull_janssen_oerding_99}. 

Other values of $r$ are related to the fractal dimension of substructures of 
percolation clusters. Consider $r \to -1^+$. One obtains immediately as a 
consequence of Eq.~(\ref{powerInTermsOfI}), that
\begin{eqnarray}
\label{rTo-1}
R_{-1} (x ,x^\prime) = \lim_{r\to -1^+} \sum_{\langle i,j \rangle} \rho_{i,j}  \left| \frac{I_{i,j}}{I}\right|^{r+1} = \sum_{\langle i,j \rangle} \rho_{i,j} \ ,
\end{eqnarray}
with only those bonds carrying non zero current contributing to the sum on the right hand side. Hence 
\begin{eqnarray}
M_{-1} (x ,x^\prime) \sim M_B \ ,
\end{eqnarray}
where $M_B$ stands for the number of bonds belonging to the backbone. Thus, the 
fractal dimension of the backbone can be expressed as
\begin{eqnarray}
D_B = \lim_{r\to -1^+} \phi_r / \nu \ .
\end{eqnarray}

Now we turn to $r \to \infty$ and $r \to 0^+$ following the lines of Blumenfeld 
and Aharony\cite{blumenfeld_aharony_85}. On the backbone between two sites $x$ and $x^\prime$ one may distinguish between two different substructures: blobs formed by multi-connected bonds and singly connected bonds which are referred to as red bonds. Both substructures are contributing to the resistance of the backbone
\begin{eqnarray}
R_{r} (x ,x^\prime) = \sum_{\langle i,j \rangle}^{\mbox{\scriptsize blob \normalsize}} 
\rho_{i,j} \left| \frac{I_{i,j}}{I}\right|^{r+1} + \sum_{\langle i,j \rangle}^{\mbox{\scriptsize red \normalsize}} 
\rho_{i,j} \ ,
\end{eqnarray}
where the sums are taken over all bonds belonging to blobs and over all red bonds respectively. Since sites on a blob are multi-connected by definition, $I_{i,j} < I$, and thus
\begin{eqnarray}
\lim_{r\to \infty} \sum_{\langle i,j \rangle}^{\mbox{\scriptsize blob \normalsize}} 
\rho_{i,j} \left| \frac{I_{i,j}}{I}\right|^{r+1} = 0 
 \ .
\end{eqnarray}
In conclusion, the dimension of the red bonds is related to $\phi_r$ via
\begin{eqnarray}
d_{\mbox{{\scriptsize red}}} = \lim_{r\to \infty} \phi_r / \nu  \ .
\end{eqnarray}

Consider now the first site $x$ at some end of a blob. An entering current $I$ 
splits into currents $I_{i,x}$ flowing to nearest neighbors $i$ with
\begin{eqnarray}
\left| I_{i,x} \right| = \sigma_{i,x} \left| V_x - V_i \right|^s  \ .
\end{eqnarray}
In the limit $s \to \infty$ the ratios $|I_{i,x}| / |I_{j,x}|$ vanish whenever 
$\sigma_{i,x} \left| V_x - V_i \right|^s < \sigma_{j,x} \left| V_x - V_j 
\right|^s$. Thus, current flows only through the resistor with the largest 
$\sigma_{i,x} \left| V_x - V_i \right|^s$. This argument may be iterated through the entire blob. One identifies either a single self avoiding chain through 
which $I$ flows, with  
\begin{eqnarray}
\label{powerOfChain}
P = \sum_{\langle i,j \rangle} \rho_{i,j} |I|^{r+1} 
\end{eqnarray}
being the power dissipated on the chain, or several of such chains with 
identical power. The expression in Eq.~(\ref{powerOfChain}) is minimal for 
minimal $\sum_{\langle i,j \rangle} \rho_{i,j}$, i.e., the current chooses the shortest path through the blob and one is led to
\begin{eqnarray}
d_{\mbox{{\scriptsize min}}} = \lim_{r\to 0^+} \phi_r / \nu  
\end{eqnarray}
for the chemical length exponent. 

\subsection{Generating function}
\label{generatingFunction}
Our aim is to determine $M_r$. Hence our task is twofold: we need to solve the 
set of Kirchhoff's equations~(\ref{cirquitEquations}) and to perform the average over all configurations of the diluted lattice. It can be accomplished by 
employing the replica technique\cite{stephen_78,harris_lubensky_84}. The network is replicated $D$-fold: $V_x \to \vec{V_x} = \left( V_x^{(1)}, \ldots , V_x^{(D)} \right)$. One introduces
\begin{eqnarray}
\psi_{\vec{\lambda}}(x) = \exp \left( i \vec{\lambda} \cdot \vec{V}_x \right) \ ,
\end{eqnarray}
where $\vec{\lambda} \cdot \vec{V}_x = \sum_{\alpha} \lambda^{(\alpha )} 
V_x^{(\alpha )}$ and $\vec{\lambda} \neq \vec{0}$. One considers the correlation 
function
\begin{eqnarray}
G \left( x, x^\prime ;\vec{\lambda} \right) = \left\langle 
\psi_{\vec{\lambda}}(x)\psi_{-\vec{\lambda}}(x^\prime) 
\right\rangle_{\mbox{\scriptsize{rep}}}
\end{eqnarray} 
given by
\begin{eqnarray}
\label{erzeugendeFunktion}
G \left( x, x^\prime ;\vec{\lambda} \right) = \left\langle Z^{-D} \int \prod_j 
\prod_{\alpha =1}^D dV_j^{(\alpha )} \exp \left[ -\frac{1}{s+1} P \left( \left\{ 
\vec{V} \right\} \right) + \frac{i\omega}{2} \sum_i \vec{V}^2_i + i \vec{\lambda} \cdot \left( \vec{V}_x  - 
\vec{V}_{x^\prime} \right) \right] \right\rangle_C \ .
\end{eqnarray}
Here $P \left( \left\{ \vec{V} \right\} \right) = \sum_{\alpha =1}^D P \left( \left\{ V^{(\alpha )} \right\} \right) = \sum_{\alpha =1}^D \sum_{\langle i,j \rangle} \sigma_{i,j} 
\left| V_i^{(\alpha)} - V_j^{(\alpha)}\right|^{s+1}$ and $Z$ is the 
normalization
\begin{eqnarray}
\label{norm}
Z = \int \prod_{j} dV_{j} \exp \left[ -\frac{1}{s+1} P \left( \left\{ V \right\} \right) + \frac{i\omega}{2} \sum_i V^2_i \right] \ .
\end{eqnarray}
Note that we have introduced an additional power term $\frac{i\omega}{2} \sum_i \vec{V}^2_i$. This is necessary to give the integrals in Eqs.~(\ref{erzeugendeFunktion}) and (\ref{norm}) a well defined meaning. Without this term the integrands depend only on voltage differences and the integrals are divergent. Physically the new term corresponds to grounding each lattice site by a capacitor of unit capacity. The original situation may be restored by taking the limit of vanishing frequency, $\omega \to 0$.

In contrast to the linear network, $P$ is not quadratic and hence the 
integration over the voltages is not Gaussian. This obstacle may be surmounted 
by employing the saddle point method\cite{harris_87}. The saddle point equation is identical to the variation principle stated in Eq.~(\ref{variationPrinciple1}). Thus the maximum of the integrand is determined by the solution of the circuit equations (\ref{cirquitEquations}) and, up to an unimportant multiplicative constant which goes to one in the limit $D \to 0$,
\begin{eqnarray}
\label{GenFkt}
G \left( x, x^\prime ;\vec{\lambda} \right) = \left\langle \exp \left(  
\frac{ \Lambda_r \left( \vec{\lambda} \right)}{r+1} R_r \left( x,x^\prime \right) \right) \right\rangle_C 
\end{eqnarray}
where
\begin{eqnarray}
\Lambda_r \left( \vec{\lambda} \right)=\sum_{\alpha =1}^D \left( - i \lambda^{(\alpha )} 
\right)^{r+1} \ .
\end{eqnarray}
Consequently, $G \left( x, x^\prime ;\vec{\lambda} \right)$ 
may serve as a generating function for $M_r$, which may be obtained by taking 
the derivative of
\begin{eqnarray}
\label{expansionOfG}
G \left( x, x^\prime ;\vec{\lambda} \right) = \left\langle \chi ( x,  x^\prime ) 
\right\rangle_C \left( 1 + \frac{ \Lambda_r \left( \vec{\lambda} \right)}{r+1} M_r ( x, x^\prime ) + \ldots 
\right) \ ,
\end{eqnarray}
with respect to $\Lambda_r$ evaluated at $\vec{\lambda}^2 = 0$. 

At this point a comment on the nature of $\vec{\lambda}$ is appropriate. One sets
\begin{eqnarray}
\label{crazyLambda}
\lambda^{(\alpha )} = i \lambda_0 + \xi^{(\alpha )} \ ,
\end{eqnarray}
with real positive $\lambda_0$ and $\xi^{(\alpha )}$ and imposes the condition $\sum_{\alpha =1}^D 
\xi^{(\alpha )} = 0$. The saddle point approximation in Eq.~(\ref{GenFkt}) may 
be justified by demanding
\begin{eqnarray}
\label{cond1}
\lambda_0 \gg 1 . 
\end{eqnarray}
On the other hand, substitution of Eq.~(\ref{crazyLambda}) into the definition 
of $\Lambda_r$ leads to
\begin{eqnarray}
 \Lambda_r \left( \vec{\lambda} \right) &=& \sum_{\alpha =1}^D \left\{ \lambda_0^{r+1} - i \left( r+1 \right) 
\lambda_0^{r} \xi^{(\alpha )} - \frac{r \left( r+1 \right) }{2} \lambda_0^{r-1} 
\xi^{(\alpha ) 2} + \ldots \right\}
\nonumber \\
&=&
D  \lambda_0^{r+1} - \frac{r \left( r+1 \right) }{2} \lambda_0^{r-1} \vec{\xi}^2 
+ \ldots \ .
\end{eqnarray}
Thus one can justify the expansion in Eq.~(\ref{expansionOfG}) by invoking the 
conditions
\begin{eqnarray}
\label{cond2}
\lambda_0^{r+1} \ll D^{-1} \quad \mbox{and} \quad \lambda_0^{r-1} \vec{\xi}^2 
\ll 1 \ .
\end{eqnarray}
Note that the replica limit $D\to 0$ allows for a simultaneous fulfilment of the conditions (\ref{cond1}) and (\ref{cond2}). However, we will not only rely on these conditions on $\vec{\lambda}$. We will provide several consistency checks for the validity of Harris' saddle point approach as we go along and reproduce known results.

\subsection{Field theoretic Hamiltonian}
\label{fieldTheoreticHamiltonian}
Since infinite voltage drops between different clusters may occur, it is not guaranteed that $Z$ stays finite, i.e., the limit $\lim_{D\to 0}Z^D$ is not well defined. Moreover, $\vec{\lambda} = \vec{0}$ has to be excluded properly. Both problems can be handled by restoring to a lattice regularization of the integrals in Eqs.~(\ref{erzeugendeFunktion}) and (\ref{norm}). One switches to voltage variables $\vec{\theta}= \Delta \theta \vec{k}$ taking 
discrete values on a $D$-dimensional torus, i.e.\ $\vec{k}$ is chosen to be an 
$D$-dimensional integer with $-M < k^{(\alpha)} \leq M$ and $k^{(\alpha 
)}=k^{(\alpha )} \mbox{mod} (2M)$. In this discrete picture there are $(2M)^D-1$ independent state variables per lattice site and one can introduce the Potts spins\cite{Zia_Wallace_75} 
\begin{eqnarray}
\Phi_{\vec{\theta}} \left( x \right) = (2M)^{-D} \sum_{\vec{\lambda} \neq 
\vec{0}} \exp \left( i \vec{\lambda} \cdot \vec{\theta} \right) 
\psi_{\vec{\lambda}} (x) = \delta_{\vec{\theta}, \vec{\theta}_{x}} - (2M)^{-D} 
\end{eqnarray}
subject to the condition $\sum_{\vec{\theta }} \Phi_{\vec{\theta}} \left( x 
\right) = 0$.

Now we revisit Eq.~(\ref{erzeugendeFunktion}). Carrying out the average over the diluted lattice configurations provides us with the weight $\exp 
(-H_{\mbox{\scriptsize{rep}}})$ of the average $\langle \cdots 
\rangle_{\mbox{\scriptsize{rep}}}$,
\begin{eqnarray}
H_{\mbox{\scriptsize{rep}}} &=&  - \ln \left\langle  \exp \left( - \frac{1}{s+1} P \left( \vec{\theta} \right) + \frac{i\omega}{2} \sum_i \vec{\theta}_{i}^2 \right) \right\rangle_C 
\nonumber \\
&=& - \sum_{\langle i ,j \rangle} \ln \left\langle  \exp \left( - 
\frac{1}{s+1} \sigma_{i ,j} \left| \theta_{i} - \theta_{j} 
\right|^{s+1} \right) \right\rangle_C + \frac{i\omega}{2} \sum_i \vec{\theta}_{i}^2 \ ,
\end{eqnarray}
where we have introduced the abbreviation $\left| \theta \right|^{s+1} = \sum_{\alpha = 1}^D \left| \theta^{(\alpha )} \right|^{s+1}$.
By dropping a constant term $N_B \ln (1-p)$, with $N_B$ being the number of 
bonds in the undiluted lattice, one obtains
\begin{eqnarray}
H_{\mbox{\scriptsize{rep}}} &=& - \sum_{\langle i ,j \rangle} K \left( \vec{\theta}_{i} - \vec{\theta}_{j} \right) + \sum_i h \left( \vec{\theta}_{i} \right)
\nonumber \\
&=& - \sum_{\langle i ,j \rangle} \sum_{\vec{\theta},\vec{\theta}^\prime} K \left( \vec{\theta} - \vec{\theta}^\prime \right) \Phi_{\vec{\theta}} \left( i \right) \Phi_{\vec{\theta}^\prime} \left( j \right) + \sum_i \sum_{\vec{\theta}} h \left( \vec{\theta} \right) \Phi_{\vec{\theta}} \left( i \right) \ ,
\end{eqnarray}
where
\begin{eqnarray}
K \left( \vec{\theta} \right) = \ln \left\{ 1 + \frac{p}{1-p} \exp \left( - 
\frac{1}{s+1} \sigma \left| \theta \right|^{s+1} \right) \right\}
\end{eqnarray}
and
\begin{eqnarray}
h \left( \vec{\theta} \right) = \frac{i\omega}{2} \sum_i \vec{\theta}_{i}^2 \ .
\end{eqnarray}
Note that $K \left( \vec{\theta} \right)$ is an exponentially decreasing function in replica space with a decay rate proportional to $\sigma^{-1}$. For large $\sigma$, the Hamiltonian $H_{\mbox{\scriptsize{rep}}}$ describes a translationally invariant short range interaction of Potts spins in real and replica space with an external one site potential $ h \left( \vec{\theta} \right)$. Moreover, the interaction potential $K \left( \vec{\theta} \right)$ is an analytic function of $\sum_{\alpha = 1}^D \left| \theta^{(\alpha )} \right|^{s+1}$. Thus the Fourier transform
\begin{eqnarray}
\tilde{K} \left( \vec{\lambda} \right) = \frac{1}{(2M)^D} \sum_{\vec{\theta}} 
\exp \left( -i \vec{\lambda} \cdot \vec{\theta} \right) K \left( \vec{\theta} 
\right)
\end{eqnarray}
can be Taylor expanded as 
\begin{eqnarray}
\label{taylorExp}
\tilde{K} \left( \vec{\lambda} \right) = w_0 - \sum_{p=1}^{\infty} w_{r,p} \left[ - \Lambda_r \left( \vec{\lambda} \right) \right]^p \ ,
\end{eqnarray}
with $w_0$ and $w_{r,p} \sim \sigma^{-p}$ being expansion coefficients.

In the limit of perfect transport, $\sigma \to \infty$, $K \left( \vec{\theta} \right)$ goes to its local limit $K \left( \vec{\theta} \right) = K \delta_{\vec{\theta}, \vec{0}}$, with $K$ being a constant. The interaction part of the Hamiltonian reduces to
\begin{eqnarray}
\label{interActionHamil}
H^{\mbox{\scriptsize{int}}}_{\mbox{\scriptsize{rep}}} = - K \sum_{\langle i ,j \rangle}  \sum_{\vec{\theta}} \Phi_{\vec{\theta}} \left( i \right) \Phi_{\vec{\theta}} \left( j \right) \ .
\end{eqnarray}
This represents nothing more than the $\left( 2M \right)^D$ states Potts model which is invariant against all $\left( 2M \right)^D !$ permutations of the Potts spins $\Phi_{\vec{\theta}}$. If $\sigma^{-1} \neq 0$, this $S_{\left( 2M \right)^D}$ symmetry is lost in favor of the short range interaction.

We proceed with the usual coarse graining step and replace the Potts spins 
$\Phi_{\vec{\theta}} \left( x \right)$ by order parameter fields $\varphi 
\left( {\rm{\bf x}} ,\vec{\theta} \right)$ which inherit the constraint $\sum_{\vec{\theta}} \varphi \left( {\rm{\bf x}} ,\vec{\theta} \right) = 0$. We model
the corresponding field theoretic Hamiltonian $\mathcal H \mathnormal$ in the spirit of Landau as a mesoscopic free energy from local monomials of the order parameter field and its gradients in real and replica space. The gradient expansion is justified since the interaction is short ranged in both spaces. Purely local terms in replica space have to respect the full $S_{\left( 2M \right)^D}$ Potts symmetry. After these remarks we write down the Landau-Ginzburg-Wilson type Hamiltonian
\begin{eqnarray}
\label{finalHamil}
\mathcal{H} &=& \int d^dx \sum_{\vec{\theta}} \Bigg\{ \frac{\tau}{2} \varphi 
\left( {\rm{\bf x}} , \vec{\theta} \right)^2 - \frac{w_r}{2} \varphi \left( 
{\rm{\bf x}} , \vec{\theta} \right) \sum_{\alpha =1}^D \left( - \frac{\partial}{\partial \theta^{(\alpha )}} \right)^{r+1} \varphi \left( {\rm{\bf x}} , \vec{\theta} \right) 
\nonumber \\
&+&  \frac{1}{2} \left( \nabla \varphi \left( {\rm{\bf x}} , \vec{\theta} \right) \right)^2 + \frac{g}{6}\varphi \left( {\rm{\bf x}} , \vec{\theta} 
\right)^3 + \frac{i \omega}{2} \vec{\theta}^2 \varphi \left( {\rm{\bf x}} , \vec{\theta} \right) \Bigg\} \ .
\end{eqnarray}
Here we have neglected all terms that are irrelevant in the renormalization group sense. $\tau$ and $w_r$ are now coarse grained analogues of the original coefficients $w_0$ and $w_{r,1}$ appearing in Eq.~(\ref{taylorExp}). Terms associated with $w_{r,p}$ are irrelevant for $p\geq 2$ and therefore neglected. Note again that $\mathcal H \mathnormal$ reduces to the usual $(2M)^D$ states Potts model Hamiltonian by setting $w_r =0$ as one retrieves purely geometrical percolation in the limit of vanishing $w_r$ ($\sigma \to \infty$). 

\section{Renormalization Group Analyses}
\label{RGA}

\subsection{Resistance of Feynman Diagrams}
\label{resistanceOfFeynmanDiagrams}
The diagrammatic elements contributing to our renormalization group improved perturbation calculation are the three point vertex $-g$ and the propagator
\begin{eqnarray}
\label{propagatorDecomp}
\frac{ 1 - \delta_{\vec{\lambda}, \vec{0}}}{{\rm{\bf p}}^2 + \tau - w_r \Lambda_r\left( \vec{\lambda} \right)} 
= \frac{1}{{\rm{\bf p}}^2 + \tau - w_r \Lambda_r \left( \vec{\lambda} \right)} - \frac{\delta_{\vec{\lambda}, 
\vec{0}}}{{\rm{\bf p}}^2 + \tau} \ .
\end{eqnarray}
Note that we have switched to a $\left( {\rm{\bf p}} , \vec{\lambda} \right)$-representation by employing Fourier transformation in real and replica space. Eq.~(\ref{propagatorDecomp}) shows that the principal propagator decomposes into a propagator carrying $\vec{\lambda}$'s (conducting) and one not carrying 
$\vec{\lambda}$'s (insulating). This allows for a schematic decomposition of 
principal diagrams into sums of diagrams consisting of conducting and insulating propagators (see App.~\ref{app:decomposition}). Here a new interpretation of the Feynman diagrams emerges\cite{stenull_janssen_oerding_99}. They may be viewed as resistor networks themselves with conducting propagators corresponding to conductors and insulating propagators corresponding to open bonds. The 
parameters $s$ appearing in a Schwinger parametrization of the conducting 
propagators,
\begin{eqnarray}
\frac{1}{{\rm{\bf p}}^2 + \tau - w_r \Lambda_r \left( \vec{\lambda} \right)} = \int_0^\infty ds \exp \left[ - 
s \left( \tau + {\rm{\bf p}}^2 - w_r {\Lambda_r \left( \vec{\lambda} \right)} \right) \right]  \ ,
\end{eqnarray}
correspond to resistances and the replica variables $i\vec{\lambda}$ to currents. The replica currents are conserved in each vertex and we may write for each edge $i$ of a diagram, $\vec{\lambda}_i = \vec{\lambda}_i \left( \vec{\lambda} , \left\{ \vec{\kappa} \right\} \right)$, where $\vec{\lambda}$ is an external current and $\left\{ \vec{\kappa} \right\}$ denotes a complete set of independent loop currents. The $\vec{\lambda}$-dependent part of a diagram can be expressed in terms of its power $P$:
\begin{eqnarray}
\exp \left( w_r \sum_{i} s_i \Lambda_r \left( \vec{\lambda}_i \right) \right) = \exp \left[ w_r P \left( 
\vec{\lambda} , \left\{ \vec{\kappa} \right\} \right) \right] \ .
\end{eqnarray} 

The new interpretation suggests an alternative way of computing the Feynman 
diagrams. To evaluate sums over independent loop currents
\begin{eqnarray}
\label{toEvaluate}
\sum_{\left\{ \vec{\kappa} \right\}} \exp \left[ w_r P \left( \vec{\lambda} , 
\left\{ \vec{\kappa} \right\} \right) \right] 
\end{eqnarray}
we employ the saddle point method under the conditions discussed at the end of Sec.~\ref{generatingFunction}. Note that the saddle point equation is 
nothing more than the variation principle stated in 
Eq.~(\ref{variationPrinciple2}). Thus solving the saddle point equations is 
equivalent to determining the total resistance $R \left( \left\{ s_i \right\} 
\right)$ of a diagram, and the saddle point evaluation of (\ref{toEvaluate}) 
yields
\begin{eqnarray}
\exp \left[ R_r \left(  \left\{ s_i \right\} \right) w_r \Lambda_r \left( \vec{\lambda} \right) \right] \ , 
\end{eqnarray}
where we have omitted once more multiplicative factors which go to one for $D \to0$. A completion of squares in the momenta renders the momentum integrations 
straightforward. Equally well we can use the saddle point method which is exact here since the momentum dependence is purely quadratic. After an expansion for small $\Lambda_r \left( \vec{\lambda} \right)$ all diagrammatic contributions are of the form
\begin{eqnarray}
\label{expansionOfDiagrams}
I \left( {\rm{\bf p}}^2 , \vec{\lambda}^2 \right) &=& I_P \left( {\rm{\bf p}}^2 \right) + I_W \left( {\rm{\bf p}}^2 \right) w_r \Lambda_r \left( \vec{\lambda} \right) + \ldots
\nonumber \\
&=& \int_0^\infty \prod_i ds_i \left[ 1 + R_r \left(  \left\{ s_i \right\} 
\right) w_r \Lambda_r \left( \vec{\lambda} \right) + \ldots \right] D \left( {\rm{\bf p}}^2, \left\{ s_i \right\} \right) \ .
\end{eqnarray}
$D \left( {\rm{\bf p}}^2, \left\{ s_i \right\} \right)$ is nothing more than the integrand one obtains upon Schwinger parameterization of the corresponding diagram in the usual $\phi^3$ theory.

\subsection{Renormalization and scaling}
\label{renormalizationAndScaling}
We proceed with standard techniques of renormalized field theory\cite{amit_zinn-justin}. The ultraviolet divergences occurring in the diagrams can be regularized by dimensional regularization. We employ the 
renormalization scheme
\begin{mathletters}
\begin{eqnarray}
\psi \to {\mathaccent"7017 \psi} = Z^{1/2} \psi \ ,&\quad\quad&
\tau \to {\mathaccent"7017 \tau} = Z^{-1} Z_{\tau} \tau \ ,
 \\
w_r \to {\mathaccent"7017 w}_r = Z^{-1} Z_{w_r} w_r \ , &\quad&
g \to {\mathaccent"7017 g} = Z^{-3/2} Z_u^{1/2} G_\epsilon^{-1/2} u^{1/2} 
\mu^{\epsilon /2} \ ,
\end{eqnarray}
\end{mathletters}
where $\epsilon = 6-d$ and $\mu$ is an inverse length scale. The factor 
$G_\epsilon = (4\pi )^{-d/2}\Gamma (1 + \epsilon /2)$, with $\Gamma$ denoting 
the Gamma function, is introduced for convenience. The Z factors may be determined by minimal subtraction, i.e., they are chosen to solely cancel poles in $\epsilon$. $Z$, $Z_\tau$ and $Z_u$ are the usual Potts model $Z$ factors. They have been computed to three loop order by be Alcantara Bonfim {\it et al}\cite{alcantara_80}. It remains to calculate $Z_{w_r}$. We postpone this calculation to Sec.~\ref{fractalDimensions}.

The unrenormalized theory has to be independent of the length scale $\mu^{-1}$ 
introduced by renormalization. In particular, the connected $N$ point 
correlation functions must be independent of $\mu$, i.e.,  
\begin{eqnarray}
\label{independence}
\mu \frac{\partial}{\partial \mu} {\mathaccent"7017 G}_N \left( \left\{ {\rm{\bf 
p}}, {\mathaccent"7017 w}_r \Lambda_r \left( \vec{\lambda} \right)\right\} ; {\mathaccent"7017 \tau}, {\mathaccent"7017 g} \right) = 0
\end{eqnarray}
for all $N$. Eq.~(\ref{independence}) translates via the Wilson functions
\begin{mathletters}
\begin{eqnarray}
\label{wilson}
\beta \left( u \right) = \mu \frac{\partial u}{\partial \mu} \bigg|_0 \ ,
&\quad& 
\kappa \left( u \right) = \mu \frac{\partial
\ln \tau}{\partial \mu}  \bigg|_0 \ ,
 \\
\zeta_r \left( u \right) = \mu \frac{\partial \ln w_r}{\partial \mu}  \bigg|_0 \ ,
&\quad& 
\gamma_{...} \left( u \right) = \mu \frac{\partial }{\partial \mu} \ln Z_{...}  \bigg|_0 \ ,
\end{eqnarray}
\end{mathletters}
where the bare quantities are kept fix while taking the derivatives, into the 
Gell-Mann-Low renormalization group equation
\begin{eqnarray}
\left[ \mu \frac{\partial }{\partial \mu} + \beta \frac{\partial }{\partial u} + 
\tau \kappa \frac{\partial }{\partial \tau} + w_r \zeta_r \frac{\partial }{\partial w_r} + \frac{N}{2} \gamma \right] G_N \left( \left\{ {\rm{\bf x}} ,w_r \Lambda_r \left( \vec{\lambda} \right) 
\right\} ; \tau, u, \mu \right) = 0 \ .
\end{eqnarray}
The particular form of the Wilson functions can be extracted from the 
renormalization scheme and the $Z$ factors.

The renormalization group equation is solved by the method of characteristics. At the infrared stable fixed point $u^\ast$, determined by $\beta \left( u^\ast \right) = 0$, the solution reads
\begin{eqnarray}
\label{SolOfRgg}
G_N \left( \left\{ {\rm{\bf x}} ,w_r \Lambda_r \left( \vec{\lambda} \right) \right\} ; \tau, u, \mu \right) = 
l^{\gamma^\ast N/2} G_N \left( \left\{ l{\rm{\bf x}} ,l^{\zeta_r^\ast}w_r \Lambda_r \left( \vec{\lambda} \right) \right\} ; l^{\kappa^\ast}\tau , u^\ast, l \mu \right) \ , 
\end{eqnarray}
where $\gamma^\ast = \gamma \left( u^\ast \right)$, $\kappa^\ast = \kappa \left( u^\ast \right)$ and $\zeta_r^\ast = \zeta_r \left( u^\ast \right)$.

To get a scaling relation for the correlation functions, a dimensional analysis remains to be performed. It yields
\begin{eqnarray}
\label{dimAna}
G_N \left( \left\{ {\rm{\bf x}} ,w_r \Lambda_r \left( \vec{\lambda} \right) \right\} ; \tau, u, \mu \right) = 
\mu^{(d-2) N/2} G_N \left( \left\{ \mu {\rm{\bf x}} ,\mu^{-2}w_r \Lambda_r \left( \vec{\lambda} \right) \right\} ; \mu^{-2}\tau , u, 1 \right) \ . 
\end{eqnarray}
From Eqs.~(\ref{SolOfRgg}) and (\ref{dimAna}) we drive the scaling relation
\begin{eqnarray}
\label{scaling}
G_N \left( \left\{ {\rm{\bf x}} ,w_r \Lambda_r \left( \vec{\lambda} \right) \right\} ; \tau, u, \mu \right) = 
l^{(d-2+\eta)N/2} G_N \left( \left\{ l{\rm{\bf x}} , l^{-\phi_r/\nu}w_r \Lambda_r \left( \vec{\lambda} \right) \right\} ; 
l^{-1/\nu}\tau , u^\ast, \mu \right) \ ,
\end{eqnarray}
with the well known critical exponents for percolation\cite{alcantara_80}
\begin{eqnarray}
\label{eta}
\eta = \gamma^\ast = - \frac{1}{21}\epsilon - 
\frac{206}{9261}\epsilon^2 + \left[ - \frac{93619}{8168202} + \frac{256}{7203} 
\zeta \left( 3 \right) \right]\epsilon^3 + {\sl O} \left( \epsilon^4  \right) \ ,
\end{eqnarray}
and
\begin{eqnarray}
\nu = \left( 2 - \kappa^\ast \right)^{-1} = \frac{1}{2} + 
\frac{5}{84}\epsilon + \frac{589}{37044}\epsilon^2 + \left[ 
\frac{716519}{130691232} - \frac{89}{7203} \zeta \left( 3 \right) 
\right]\epsilon^3 + {\sl O} \left( \epsilon^4  \right) \ .
\end{eqnarray}
Note that $\zeta$ in Eq.~(\ref{eta}) stands for the Riemann zeta function and 
should not be confused with the Wilson function defined above.
The exponent $\phi_r$ is defined by 
\begin{eqnarray}
\phi_r = \nu \left( 2 - \zeta_r^\ast \right) = \nu \left( 2 - 
\eta + \psi_r \right)
\end{eqnarray}
with $\psi_r = \gamma_{w_r} \left( u^\ast \right)$. It has been calculated for arbitrary $r$ to one-loop order by Harris\cite{harris_87}. He found
\begin{eqnarray}
\phi_r = 1 + \frac{\epsilon}{14} c_r + {\sl O} \left( \epsilon^2  \right) \ ,
\end{eqnarray}
where
\begin{eqnarray}
c_r = \frac{1}{2} \int_{-1}^1 d\xi \frac{\left( 1 - \xi^2 \right)}{\left[ \left( 1 + \xi \right)^{1/r} + \left( 1 - \xi \right)^{1/r} \right]^r} \ .
\end{eqnarray}
Our calculation gives the same result.

Equation~(\ref{scaling}) implies the following scaling behavior of the two point correlation function $G=G_2$ at criticality,
\begin{eqnarray}
\label{scaleRel}
G \left( |{\bf x}-{\bf x}^\prime |; w_r  \Lambda_r \left( \vec{\lambda} \right) \right) = l^{d-2+\eta} G \left( l |{\bf x}-{\bf x}^\prime|; l^{-\phi_r/\nu} w_r \Lambda_r \left( \vec{\lambda} \right) \right) \ ,
\end{eqnarray}
where we dropped several arguments for notational simplicity. The choice $l = 
|{\bf x}-{\bf x}^\prime |^{-1}$ and a Taylor expansion of the right hand side of Eq.~(\ref{scaleRel}) lead to 
\begin{eqnarray}
G \left( |{\bf x}-{\bf x}^\prime |; w_r  \Lambda_r \left( \vec{\lambda} \right) \right) = |{\bf x}-{\bf x}^\prime |^{2-d-\eta} \left( 1 + w_r  \Lambda_r \left( \vec{\lambda} \right) |{\bf x}-{\bf x}^\prime |^{\phi_r/\nu} + \ldots 
\right) \ .
\end{eqnarray}
Comparison with Eq.~(\ref{expansionOfG}) gives us the scaling behavior of the 
average resistance:
\begin{eqnarray}
\label{MbehavesAs2}
M_r ({\rm{\bf x}}, {\rm{\bf x}}^\prime) \sim |{\rm{\bf x}} - {\rm{\bf x}}^\prime |^{\phi_r / \nu} \ . 
\end{eqnarray}

\section{Fractal Dimensions}
\label{fractalDimensions}
In this section we calculate $\phi_r$ for $r \to \infty$, $r \to 0^+$ and $r \to -1^+$. As discussed in Sec.~\ref{clusterProperties}, this provides us with the fractal dimension of the red bonds, the chemical length, and the backbone 
respectively.

\subsection{Red bonds}
\label{redBonds}
Consider $r \to \infty$. As argued in Sec.~\ref{clusterProperties}, blobs do not contribute to the total resistance. Now we take direct advantage of our view of the Feynman diagrams as being resistor networks themselves. In analogy to real networks, the resistance of closed loops vanishes. Only singly connected conducting propagators contribute to the total resistance of a diagram, i.e,
\begin{eqnarray}
R_\infty \left( \left\{ s_i \right\} \right) =  \sum_i^{\mbox{\scriptsize singly}} s_i\ ,
\end{eqnarray}
with the sum being taken only over singly connected conducting propagators. The contribution of a diagram to the renormalization factor $Z_{w_\infty}$ takes the form 
\begin{eqnarray}
\label{expansionOfDiagrams_r_infty}
I_W \left( {\rm{\bf p}}^2 \right) = \int_0^\infty \prod_j ds_j \sum_i^{\mbox{\scriptsize singly}} s_i D \left( {\rm{\bf p}}^2, \left\{ s_j \right\} \right) \ .
\end{eqnarray}
Note that a factor $s_i$ in Eq.~(\ref{expansionOfDiagrams_r_infty}) corresponds to the insertion of $\frac{1}{2} \varphi^2$ into the $i$th edge of the diagram. We generate $I_W \left( {\rm{\bf p}}^2 \right)$ by inserting $\frac{1}{2} \varphi^2$ in each singly connected conducting propagator. This procedure is carried out up to three loop order, i.e., every conducting propagator in App.~\ref{app:decomposition} that does not belong to a closed loop gets an insertion. For detail see App.~\ref{app:diagramsForReds}. The resulting diagrams are displayed in Fig.~\ref{diagrammaticExpansionRedBonds}.

Now consider the contributions of the diagrams listed in App.~\ref{app:decomposition} to $Z_\tau$. These can be generated by inserting $\frac{1}{2} 
\varphi^2$ in conducting as well as in insulating propagators. Again, one 
obtains the diagrams depicted in Fig.~\ref{diagrammaticExpansionRedBonds} with the same fore-factors. Consequently, $Z_{w_\infty}$ and $Z_\tau$ are identical at least up to three-loop order. The same goes for the corresponding Wilson functions $\zeta_\infty$ and $\kappa$. From the definition of $\phi_r$ it 
follows that
\begin{eqnarray}
\label{resultForPhiInfty}
\phi_\infty = \frac{2-\zeta_\infty^\ast}{2-\kappa^\ast} = 1 + {\sl O} \left( \epsilon^4  \right) \ .
\end{eqnarray}
Note that this result is in agreement with the rigorous one by 
Coniglio   \cite{coniglio_81,coniglio_82}, who proved that $d_{\mbox{{\scriptsize 
red}}} = 1/\nu$. We rate this as an indication for the validity of Harris' saddle point approach.

\subsection{Chemical length}
\label{chemicalLenght}
In the limit $r \to 0^+$ only the shortest self avoiding path of conducting 
propagators contributes to the total resistance of a diagram. In other words, 
the total resistance has to be determined such that
\begin{eqnarray}
\sum_{\mbox{\scriptsize paths}} \ \sum_{i \in \mbox{\scriptsize path}} s_i
\end{eqnarray}
is minimal, where the first sum is taken over all self avoiding paths of 
conducting propagators connecting the external legs of a diagram.

For $r \to 0^+$ the conducting propagator reads
\begin{eqnarray}
\label{prop_rto0}
\frac{1}{\tau + {\rm{\bf p}}^2 + i w_0 \sum_{\alpha =1}^D \lambda^{(\alpha )}} = \frac{1}{\chi + i w_0 \lambda} \ ,
\end{eqnarray}
with $\chi = \tau + {\rm{\bf p}}^2$ and $\lambda = \sum_{\alpha =1}^D \lambda^{(\alpha )}$. We start with the two one-loop diagrams A and B (see, App.~\ref{app:decomposition}). The diagram A translates into
\begin{eqnarray}
\mbox{A} &=& \frac{g^2}{2} \int_{\rm{\bf p}} \int_0^\infty ds_1 ds_2 \exp \left[ s_1 \chi_1 + s_2 \chi_2\right] \exp \left[ - i w_0 \lambda \min \left( s_1 ,s_2 \right) \right]
\nonumber \\
&=& g^2 \int_{\rm{\bf p}} \int_0^\infty ds_1 ds_2 \exp \left[ s_1 \chi_1 + s_2 
\chi_2\right]  \exp \left[ - i w_0 \lambda  s_1 \right] \theta \left( s_2 - s_1 \right) \ ,
\end{eqnarray}
where $\theta$ denotes the step function and $\int_{\bf{\rm p}}$ is an 
abbreviation for $\left( 2 \pi \right)^d \int d^d p$. Diagram B reads
\begin{eqnarray}
\mbox{B} = \frac{g^2}{2} \int_{\rm{\bf p}} \int_0^\infty ds_1 ds_2 \exp \left[ 
s_1 \chi_1 + s_2 \chi_2\right] \exp \left[ - i w_0 \lambda  s_1 \right] \ ,
\end{eqnarray}
and hence,
\begin{eqnarray}
\mbox{A} - 2\mbox{B} = - g^2 \int_{\rm{\bf p}} \int_0^\infty ds_1 ds_2 \exp 
\left[ s_1 \chi_1 + s_2 \chi_2\right] \exp \left[ - i w_0 \lambda  s_1 \right] 
\theta \left( s_1 - s_2 \right) \ .
\end{eqnarray}

Now we take a short detour and present some features of the field theory of dynamical percolation as studied by one of us some time ago\cite{janssen_85}. The dynamical functional ${\mathcal J}$ that leads to the diagrammatic expansion for the calculation of correlation and response functions is given by 
\begin{eqnarray}
\label{dynFunc}
{\mathcal J} = \int d^d x dt \gamma \tilde{\varphi} \left[ \gamma^{-1} \frac{\partial}{\partial t} + \left( \tau - \Delta \right) + g \phi - \frac{g}{2} \tilde{\varphi} \right] \varphi \ .
\end{eqnarray}
Here, $\phi \left( {\rm{\bf x}}, t \right) = \gamma \int_{-\infty}^t dt^\prime \varphi \left( {\rm{\bf x}}, t^\prime  \right)$ and $\tilde{\varphi} \left( {\rm{\bf x}}, t \right)$ is the response field.
\begin{eqnarray}
G_{1,1} \left( {\rm{\bf x}}, t \right) = \left\langle \varphi \left( {\rm{\bf x}}, t \right) \tilde{\varphi} \left( {\rm{\bf 0}}, t \right) \right\rangle_{\mathcal J} 
\end{eqnarray}
is the density response function that describes a growing cluster initiated by a germ at $\left( {\rm{\bf x}}={\rm{\bf 0}}, t=0 \right)$ which percolates at the critical point. Near this percolation point the response function scales as
\begin{eqnarray}
G_{1,1} \left( {\rm{\bf x}}, t \right) = \xi^{-(d-2+\eta )} f \left( {\rm{\bf x}}/\xi, t/\xi^z \right) \ ,
\end{eqnarray}
where f is a scaling function, $\xi = \left| \tau \right|^{-\nu}$ is the correlation length, and z is the dynamic exponent given to second order in $\epsilon$ in\cite{janssen_85}.

The diagrammatic elements of dynamical percolation are the propagator
\begin{eqnarray}
G \left( {\rm{\bf p}},t \right) = \theta \left( t \right) \exp \left[ -\gamma  \left( \tau + {\rm{\bf p}}^2 \right) t \right]
\end{eqnarray}
and the vertices $\gamma g$ and $- \gamma g \theta \left( t - t^\prime \right)$. These elements are depicted in Fig.~\ref{newElements}. Note that the Fourier transformed propagator reads 
\begin{eqnarray}
\tilde{G} \left( {\rm{\bf p}},\omega \right) = \frac{1}{i\omega + \gamma  \left( \tau + {\rm{\bf p}}^2 \right)}
\end{eqnarray}
and can be identified with (\ref{prop_rto0}) up to a factor $\gamma^{-1}$ by setting $\gamma w_0 \lambda = \omega$. Thus, the renormalization of $w_0$ is directly related to that of the kinetic coefficient $\gamma$. One finds that $z$ is related to the chemical length dimension by $z=d_{\mbox{{\scriptsize min}}}$.

The one loop contribution to the vertex function $\Gamma_{1,1} \left( {\rm{\bf p}},\omega \right)= 1/\tilde{G}_{1,1} \left( {\rm{\bf p}}, \omega \right)$ is visualized in Fig.~\ref{dynamicOneLoop}. We find
\begin{eqnarray}
\mbox{a} &=& - \left( \gamma g \right)^2 \int_{\rm{\bf p}} \int_0^\infty dt_1 dt_2 \exp \left[ \gamma \left( t_1 \chi_1 + t_2 \chi_2 \right) \right] \theta \left( t_1 - t_2 \right)  \exp \left[ - i \omega t_1 \right] \nonumber \\
&=& \mbox{A} - 2\mbox{B}
\end{eqnarray}
if we identify $\gamma t_i = s_i $.

Now we turn to the two-loop diagrams. In the same manner as in the one-loop case we obtain
\begin{eqnarray}
\lefteqn{ \mbox{C} - 4\mbox{D} - \mbox{E} + 2\mbox{F} + 4\mbox{G} }
\nonumber \\
&=& g^4 \int_{\rm{\bf p}} \int_{\rm{\bf q}} \int_0^\infty \prod_{i=1}^{5} ds_i 
\exp \left( \sum_{i=1}^{5} s_i \chi_i \right) 
\nonumber \\
&\times& \bigg\{ \exp \left[ -iw_0 \lambda \left( s_1 + s_4 + s_5 \right) \right]
 \Big\{ \theta \left( s_2 - s_1 - s_5 \right) \theta \left( s_3 - s_4 - s_5 \right)
\nonumber \\
&-& \theta \left( s_2 - s_1 - s_5 \right) - \theta \left( s_3 - s_4 - s_5 \right) + 1 \Big\}
\nonumber \\
&+& \exp \left[ -iw_0 \lambda \left( s_1 + s_3 \right) \right] \Big\{ \theta \left( s_2 + s_4 - s_1 - s_3 \right) \theta \left( s_4 + s_5 - s_3 \right) \theta \left( s_2 + s_5 - s_1 \right) 
\nonumber \\
&-& \theta \left( s_2 + s_4 - s_1 - s_3 \right) - \theta \left( s_4 + s_5 - s_3 \right) - \theta \left( s_2 + s_5 - s_1 \right) + 2 \Big\} \bigg\}
\nonumber \\
&=& \left( \gamma g \right)^4 \int_{\rm{\bf p}} \int_{\rm{\bf q}} \int_0^\infty \prod_{i=1}^{5} dt_i 
\exp \left( \sum_{i=1}^{5} t_i\gamma \chi_i \right) 
\nonumber \\
&\times& \bigg\{ \exp \left[ -i\omega \left( t_1 + t_4 + t_5 \right) \right] 
\theta \left( t_1 + t_5 - t_2 \right) \theta \left( t_4 + t_5 - t_3 \right) 
\nonumber \\
&+& \exp \left[ -i\omega \left( t_1 + t_3 \right) \right] \Big\{ \theta \left( t_1 - t_2 - t_5 \right) \theta \left( t_3 + t_5 - t_4 \right) 
\nonumber \\
&+& \theta \left( t_1 + t_3 - t_2 - t_4 \right) \left[  \theta \left( t_4 - t_3 - t_5 
\right) + \theta \left( t_3 - t_4 - t_5 \right) \right] \Big\} \bigg\}
\nonumber \\
&=& \mbox{b} + \mbox{c} + \mbox{d} + \mbox{e} \ .
\end{eqnarray}
The diagrams b, c, d, and e are depicted in Fig.~\ref{dynamicTwoLoop1}. For the second bold two-loop diagram we find
\begin{eqnarray}
\lefteqn{ \mbox{H} - \mbox{I} - 2\mbox{J} + 2\mbox{K} + \mbox{L} }
\nonumber \\
&=& g^4 \int_{\rm{\bf p}} \int_{\rm{\bf q}} \int_0^\infty \prod_{i=1}^{5} ds_i 
\exp \left( \sum_{i=1}^{5} s_i \chi_i \right) 
\nonumber \\
&\times& \bigg\{ \exp \left[ -iw_0 \lambda \left( s_1 + s_2 + s_3 \right) \right]
\Big\{ \theta \left( s_5 - s_1 - s_2 - s_3 \right) \theta \left( s_4 - s_2 \right)
\nonumber \\
&-& \theta \left( s_5 - s_1 - s_2 - s_3 \right) - \theta \left( s_4 - s_2 \right) + 1 \Big\}
\nonumber \\
&+& \exp \left[ -iw_0 \lambda s_5 \right] \frac{1}{2} \Big\{ \theta \left( s_1 + s_2 + s_3 - s_5 \right) \theta \left( s_1 + s_4 + s_3 - s_5 \right)
\nonumber \\
&-& \theta \left( s_1 + s_2 + s_3 - s_5 \right) - \theta \left( s_1 + s_4 + s_3 - s_5 \right) +1  \Big\} \bigg\}
\nonumber \\
&=& \left( \gamma g\right)^4 \int_{\rm{\bf p}} \int_{\rm{\bf q}} \int_0^\infty \prod_{i=1}^{5} dt_i 
\exp \left( \sum_{i=1}^{5} t_i \gamma \chi_i \right) 
\nonumber \\
&\times& \bigg\{ \exp \left[ -i\omega \left( t_1 + t_2 + t_3 \right) \right] 
\theta \left( t_2 - t_4 \right) \theta \left( t_1 + t_2 + t_3 - t_5 \right) 
\nonumber \\
&+& \exp \left[ -i\omega t_5 \right] \theta \left( t_5 - t_1 - t_2 - t_3 
\right) \theta \left( t_2 - t_4 \right) \bigg\}
\nonumber \\
&=& \mbox{f} + \mbox{g} \ .
\end{eqnarray}
The diagrams f and g are shown in Fig.~\ref{dynamicTwoLoop2}.

The dynamic diagrams lead to the result for the dynamic exponent $z$ stated in\cite{janssen_85}. Since we identified the two diagrammatic expansions up to 
two-loop order, the RRN gives the same result for the chemical length dimension as the dynamic approach in\cite{janssen_85}, 
\begin{eqnarray}
\label{resultForDmin}
d_{\mbox{{\scriptsize 
min}}} = 2 - \frac{\epsilon}{6} - \left[ \frac{937}{588} + \frac{45}{49} \left( 
\ln 2 - \frac{9}{10} \ln 3 
\right)\right] \left( \frac{\epsilon}{6} \right)^2 + {\sl O} \left( 
\epsilon^3  \right) \ .
\end{eqnarray}
Moreover, another consistency check for the saddle point approximation is 
fulfilled.

Obviously, $d_{\mbox{{\scriptsize min}}}$ has to approach one for $d \to 1$. 
This feature can be incorporated by a rational approximation yielding
\begin{eqnarray}
\label{rationalApproxForDmin}
d_{\mbox{{\scriptsize min}}} \approx 1 + \left( 1 - \frac{\epsilon}{5} \right) 
\left( 1 + \frac{\epsilon}{30} - 0.0301 \epsilon^2 \right) \ .
\end{eqnarray}
Due to the rich structure of $\eta$ in the percolation problem,
\begin{eqnarray}
\psi_{0} = - \frac{3}{14} \epsilon - \frac{365 + 140 \ln 2 - 126 \ln 3}{5488} 
\epsilon^2
\end{eqnarray} 
might be better suited for such a comparison than $d_{\mbox{{\scriptsize 
min}}}$. It is known exactly that $\psi_{0}$ vanishes in one dimension. A 
rational approximation yields
\begin{eqnarray}
\psi_{0} \approx \left( 1 - \frac{\epsilon}{5} \right) \left( - \frac{3}{14} 
\epsilon - 0.1018 \epsilon^2 \right) \ . 
\end{eqnarray}
$d_{\mbox{{\scriptsize min}}}$ and $\psi_{0}$ are compared to numerical 
simulations by Grassberger\cite{grassberger_92_99} in Fig.~\ref{dataDmin}. The rational approximants agree reasonably well with the numerical estimates at $d=3$. At $d=2$, the approximant for $d_{\mbox{{\scriptsize min}}}$ seems to be in conformity with the simulation result. However, the good agreement should be 
taken with caution. It might be accidental, since $\psi_{0} \left( d=2 \right)$ hardly agrees with the numerical value.

\subsection{Backbone}
\label{backbone}
Now we focus on the limit $r \to -1^+$. As argued in Sec.~\ref{clusterProperties}, the resistance of the backbone between two sites $x$ and $x^\prime$ is given by
\begin{eqnarray}
R_{-1} (x ,x^\prime) =\sum_{\langle i,j \rangle} \rho_{i,j} \ ,
\end{eqnarray}
with the sum running over all current carrying bonds of the underlying cluster. In analogy, the resistance of a Feynman diagram is given by
\begin{eqnarray}
R_{-1} \left( \left\{ s_i \right\} \right) =  \sum_i^{\mbox{\scriptsize cond}} s_i\ ,
\end{eqnarray}
where the sum is extending over all conducting propagators of the diagram. The contribution of a diagram to $Z_{w_{-1}}$ now takes the form 
\begin{eqnarray}
\label{expansionOfDiagrams_r_-1}
I_W \left( {\rm{\bf p}}^2 \right) = \int_0^\infty \prod_j ds_j \sum_i^{\mbox{\scriptsize cond}} s_i D \left( {\rm{\bf p}}^2, \left\{ s_j \right\} \right) \ .
\end{eqnarray}
We proceed in the same manner as in Sec.~\ref{redBonds}. However, now $\frac{1}{2} \phi^2$ is inserted into all conducting propagators. For details of the calculation see App.~\ref{app:evaluation_r_-1}. Minimal subtraction leads to the renormalization factor
\begin{eqnarray}
\label{zFactor}
Z_{w_{-1}} = 1 + \frac{u^2}{4\epsilon } + \frac{u^3}{\epsilon^2} \left[ \frac{7}{12} -\frac{29}{144}\epsilon - \frac{2}{3} \zeta \left( 3 \right) \epsilon \right] +{\sl O} \left( u^4  \right) \ .
\end{eqnarray}
Via the Wilson functions we obtain the exponents
\begin{eqnarray}
\psi_{-1} = -2 \left( \frac{\epsilon}{7} \right)^2 + \left[ 16 \zeta \left( 3 
\right) - \frac{2075}{126} \right] \left( \frac{\epsilon}{7} \right)^3 + {\sl O} \left( \epsilon^4  \right) 
\end{eqnarray}
and 
\begin{eqnarray}
D_B = 2 + \frac{1}{21} \epsilon - \frac{172}{9261} \epsilon^2 + 2 
 \frac{ - 74639 + 22680 \zeta \left( 3 \right) }{4084101} \epsilon^3 +  {\sl O}
\left( \epsilon^4  \right) \ .
\end{eqnarray}
Note that our result agrees to second order in $\epsilon$ with calculations by 
Harris and Lubensky\cite{harris_lubensky_83} based on another approach. This is again in favor of the saddle point approximation.

In Fig.~\ref{dataDb} we compare the $\epsilon$-expansions as well as the 
rational approximants
\begin{eqnarray}
\psi_{-1} \approx - \frac{2 \epsilon^2}{49} \left( 1 - \frac{\epsilon}{5} 
\right) 
\left( 1 + 1.2625 \frac{\epsilon}{500} \right) 
\end{eqnarray}
and
\begin{eqnarray}
D_B \approx 1 + \left( 1 - \frac{\epsilon}{5} \right) 
\left( 1 +  \frac{26}{105} \epsilon + \frac{7166}{231525} \epsilon^2 - 0.0170 
\epsilon^3 \right) 
\end{eqnarray}
to numerical simulations by 
Grassberger\cite{grassberger_99} and Moukarzel\cite{moukarzel_98}. At $d=4$ the results agree within the numerical errors. However, a higher accuracy of the 
numerical estimate is desirable. At $d=3$ and $d=2$ the analytic results look less realistic, but they reproduce the shape of the dependence on dimensionality.

\section{Conclusions and Outlook}
By employing a saddle point approach due to Harris we calculated the exponent $\phi_r /\nu$ for the critical behavior of the resistance in a diluted network. We focussed on distinct values of the nonlinearity $r$, namely those related to the fractal dimensions of the red bonds, the chemical path and the backbone respectively.

We provided several consistency checks for the saddle point approach. The 
validity of the approach seems to be beyond question. 

For dimensions close to the upper critical dimension six, our results for 
$d_{\mbox{\scriptsize min}}$ and $D_B$ are the most accurate analytical 
estimates that we know of. The analytic results agree reasonably well with the 
available numerical simulations. At low dimensions the agreement becomes less 
pronounced.

Our interpretation of Feynman diagrams proved to be a powerful tool. It 
simplified the renormalization group improved perturbation calculation considerably. The technique used here may be applied to other aspects of transport on percolation clusters.
For example it can be employed to calculate the family of noise exponents $\left\{ \psi_l \right\}$ for diluted resistor networks, as treated by Park, Harris and Lubensky\cite{P_H_L_87} to one-loop order. Our two-loop calculation yielding
\begin{eqnarray}
\label{monsterExponent}
\psi_l &=& 1 + \frac{\epsilon}{7  \left( 1+l \right) \left( 1+2l \right)} + \frac{\epsilon^2}{12348
      \left( 1 + l \right)^3
      \left( 1 + 2l \right)^3}
\nonumber \\
&\times& \Bigg\{
 313 - 672\gamma + 
        l\bigg\{ 3327 - 4032\gamma - 
           8l\Big\{ 4
               \left( -389 + 273\gamma
                 \right)    
\nonumber \\
&+&   
              l\left[ -2076 + 1008\gamma + 
               l  \left( -881 + 336\gamma
                     \right) \right]  \Big\}  \bigg\}  - 
        672\left( 1 + l \right)^2
         \left( 1 + 2l \right)^2
         \Psi(1 + 2l)  \Bigg\}
\end{eqnarray}
will be reported in a separate publication in the near future. In Eq.~(\ref{monsterExponent}) $\gamma$ denotes Euler's constant and $\Psi$ stands for the Digamma function.

\acknowledgements
We acknowledge support by the Sonderforschungsbereich 237 ``Unordnung und 
gro{\ss}e Fluktuationen'' of the Deutsche Forschungsgemeinschaft. O.S.\ would 
like to mention that the present work is part of a planned dissertation at the 
``Mathematisch-Naturwissenschaftliche Fakult\"{a}t der 
Heinrich-Heine-Universit\"{a}t D\"{u}sseldorf''.

\appendix
\section{Decomposition Of Diagrams}
\label{app:decomposition}
Here we list the decomposition of the primary two leg diagrams (bold) into 
conducting diagrams composed of conducting (light) and insulating (dashed) 
propagators. The listing extends up to three-loop order. Note that the conducting diagrams inherit their combinatorial factor from their bold diagram. For example, the diagrams A and B introduced below have to be calculated with the same combinatorial factor, namely $\frac{1}{2}$.
\epsfxsize=2.3cm
\begin{eqnarray}
\begin{array}{cccccc}
& \raisebox{-5mm}{\epsffile{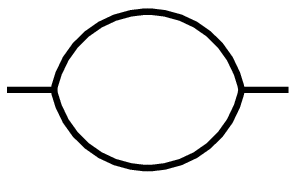}} &=& \raisebox{-5mm}{\epsffile{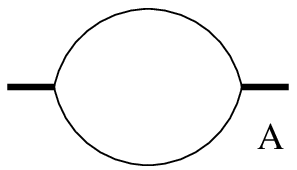}} &-2&  \raisebox{-5mm}{\epsffile{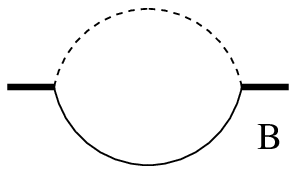}}
\end{array} 
\end{eqnarray}
\begin{eqnarray}
\begin{array}{cccccc}
& \raisebox{-5mm}{\epsffile{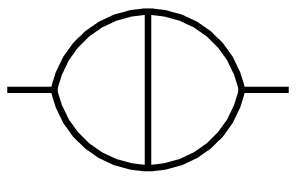}} &=& \raisebox{-5mm}{\epsffile{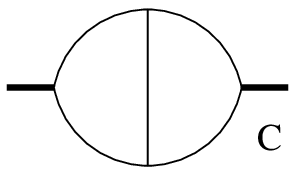}} &-4& \raisebox{-5mm}{\epsffile{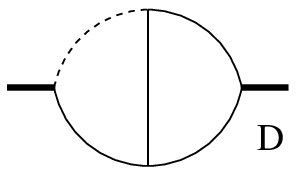}}
\vspace{2mm}
\\
-& \raisebox{-5mm}{\epsffile{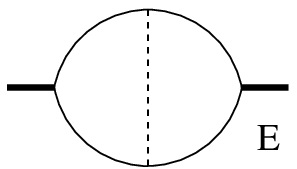}} &+2& \raisebox{-5mm}{\epsffile{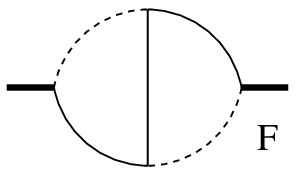}} &+4& \raisebox{-5mm}{\epsffile{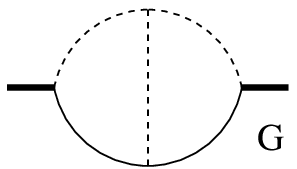}}
\end{array}
\end{eqnarray}
\begin{eqnarray}
\begin{array}{cccccc}
& \raisebox{-5mm}{\epsffile{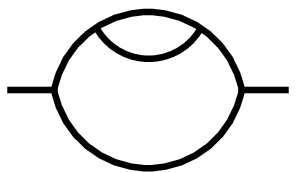}} &=& \raisebox{-5mm}{\epsffile{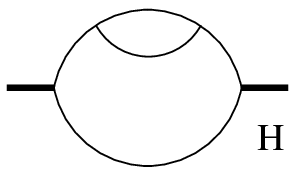}} &-& \raisebox{-5mm}{\epsffile{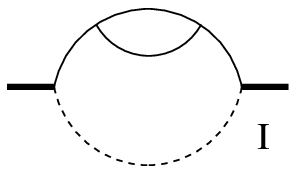}}
\vspace{2mm}
\\
-2& \raisebox{-5mm}{\epsffile{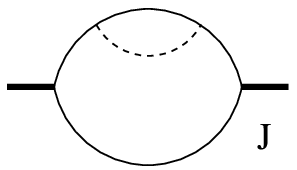}} &+2& \raisebox{-5mm}{\epsffile{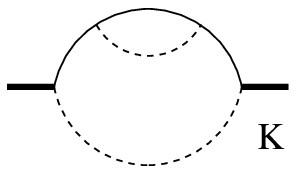}} &+& \raisebox{-5mm}{\epsffile{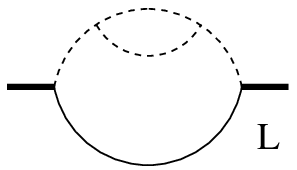}}
\end{array}
\end{eqnarray}
\begin{eqnarray}
\begin{array}{cccccc}
& \raisebox{-5mm}{\epsffile{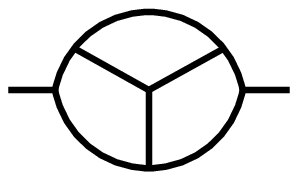}} &=& \raisebox{-5mm}{\epsffile{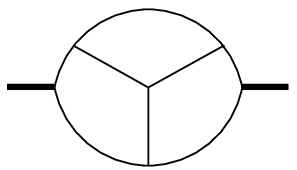}} &-2& \raisebox{-5mm}{\epsffile{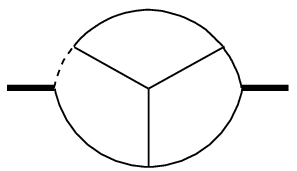}}
\vspace{2mm}
\\
-2& \raisebox{-5mm}{\epsffile{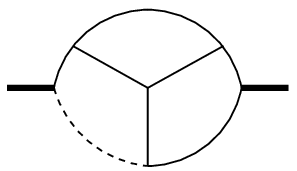}} &-2& \raisebox{-5mm}{\epsffile{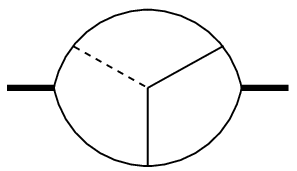}} &-& \raisebox{-5mm}{\epsffile{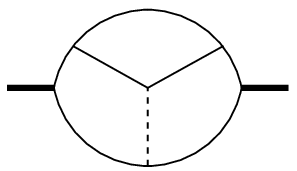}}
\vspace{2mm}
\\
-& \raisebox{-5mm}{\epsffile{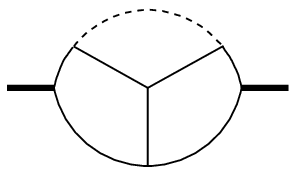}} &+2& \raisebox{-5mm}{\epsffile{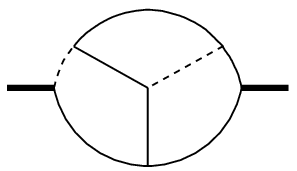}} &+2& \raisebox{-5mm}{\epsffile{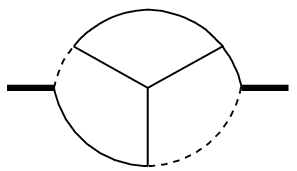}}
\vspace{2mm}
\\
+2& \raisebox{-5mm}{\epsffile{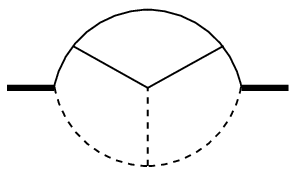}} &+4& \raisebox{-5mm}{\epsffile{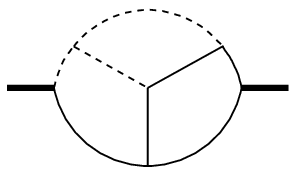}} &+2& \raisebox{-5mm}{\epsffile{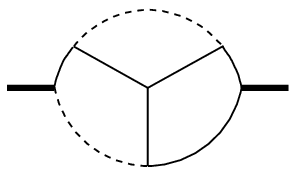}}
\vspace{2mm}
\\
+& \raisebox{-5mm}{\epsffile{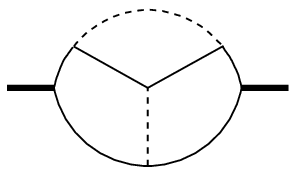}} &+2& \raisebox{-5mm}{\epsffile{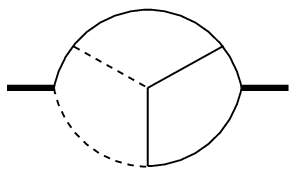}} &+2& \raisebox{-5mm}{\epsffile{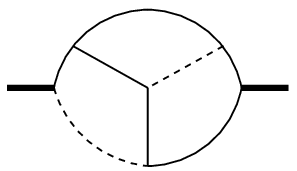}}
\vspace{2mm}
\\
+2& \raisebox{-5mm}{\epsffile{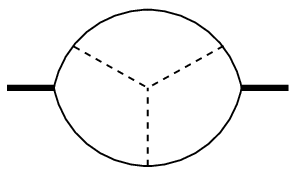}} &-4& \raisebox{-5mm}{\epsffile{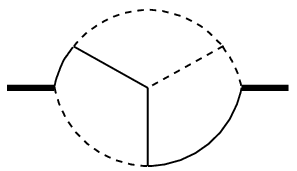}} &-2& \raisebox{-5mm}{\epsffile{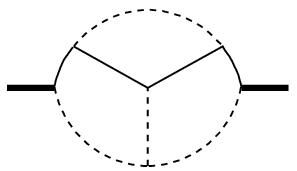}}
\vspace{2mm}
\\
-4& \raisebox{-5mm}{\epsffile{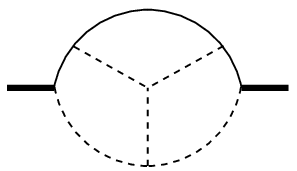}} &-4& \raisebox{-5mm}{\epsffile{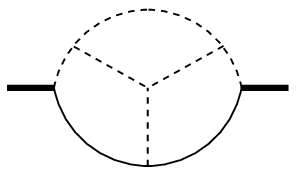}} &-2& \raisebox{-5mm}{\epsffile{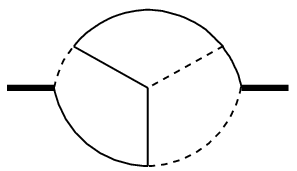}}
\end{array}
\end{eqnarray}
\begin{eqnarray}
\begin{array}{cccccc}
& \raisebox{-5mm}{\epsffile{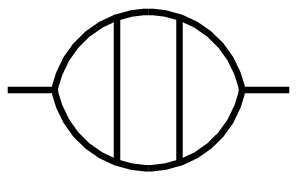}} &=& \raisebox{-5mm}{\epsffile{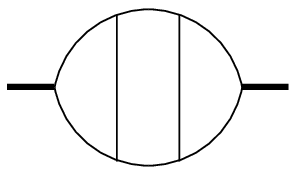}} &-4& \raisebox{-5mm}{\epsffile{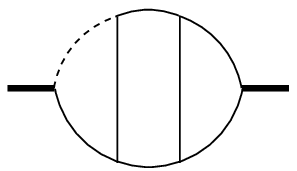}}
\vspace{2mm}
\\
-2& \raisebox{-5mm}{\epsffile{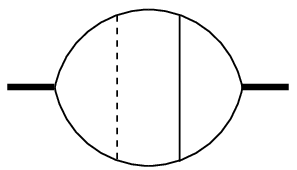}} &-2& \raisebox{-5mm}{\epsffile{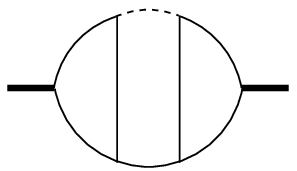}} &+2& \raisebox{-5mm}{\epsffile{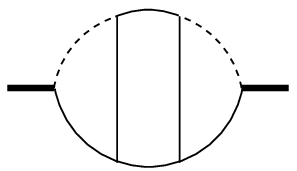}}
\vspace{2mm}
\\
+2& \raisebox{-5mm}{\epsffile{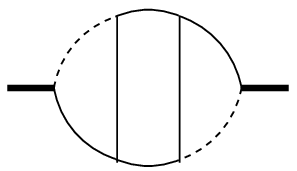}} &+4& \raisebox{-5mm}{\epsffile{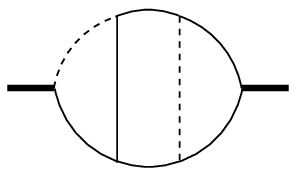}} &+8& \raisebox{-5mm}{\epsffile{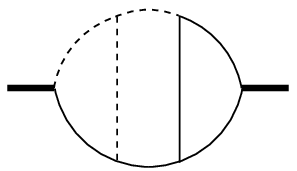}}
\vspace{2mm}
\\
+& \raisebox{-5mm}{\epsffile{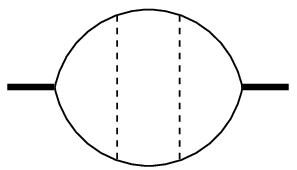}} &+4& \raisebox{-5mm}{\epsffile{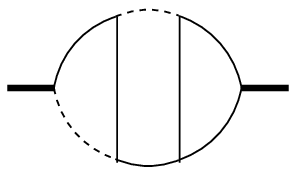}} &-8& \raisebox{-5mm}{\epsffile{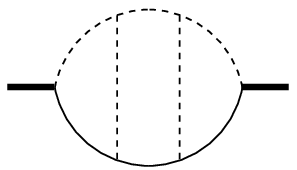}}
\vspace{2mm}
\\
-2& \raisebox{-5mm}{\epsffile{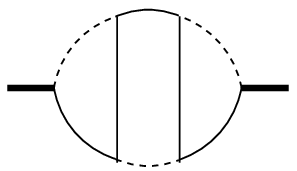}} &-8& \raisebox{-5mm}{\epsffile{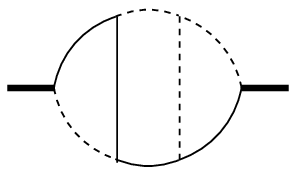}} & &
\end{array}
\end{eqnarray}
\begin{eqnarray}
\begin{array}{cccccc}
& \raisebox{-5mm}{\epsffile{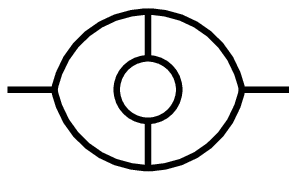}} &=& \raisebox{-5mm}{\epsffile{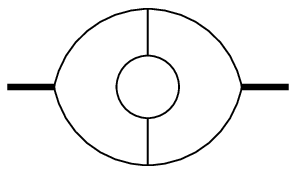}} &-4& \raisebox{-5mm}{\epsffile{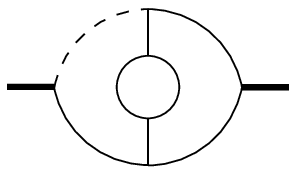}}
\vspace{2mm}
\\
-2& \raisebox{-5mm}{\epsffile{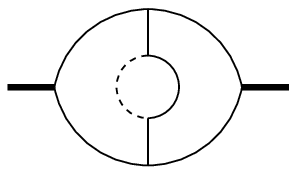}} &+& \raisebox{-5mm}{\epsffile{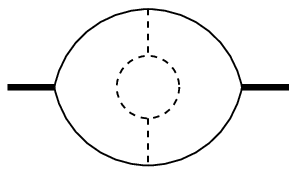}} &+2& \raisebox{-5mm}{\epsffile{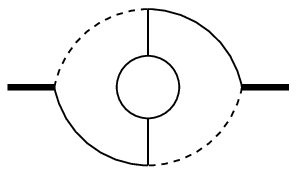}}
\vspace{2mm}
\\
+8& \raisebox{-5mm}{\epsffile{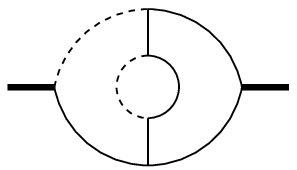}} &-4& \raisebox{-5mm}{\epsffile{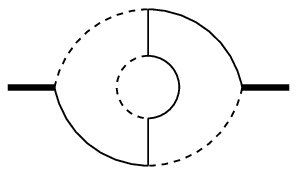}} &-4& \raisebox{-5mm}{\epsffile{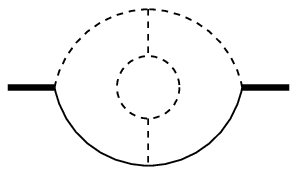}}
\end{array}
\end{eqnarray}
\begin{eqnarray}
\begin{array}{cccccc}
& \raisebox{-5mm}{\epsffile{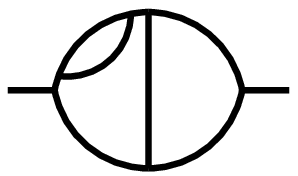}} &=& \raisebox{-5mm}{\epsffile{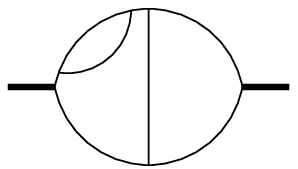}} &-& \raisebox{-5mm}{\epsffile{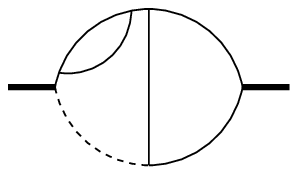}}
\vspace{2mm}
\\
-& \raisebox{-5mm}{\epsffile{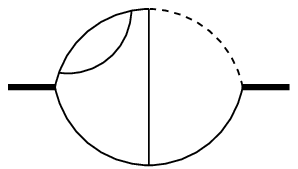}} &-& \raisebox{-5mm}{\epsffile{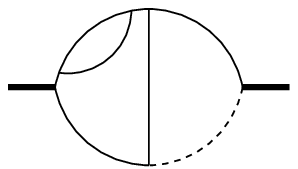}} &-& \raisebox{-5mm}{\epsffile{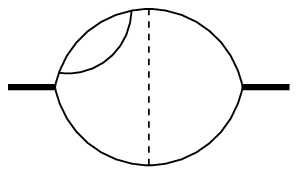}}
\vspace{2mm}
\\
-2& \raisebox{-5mm}{\epsffile{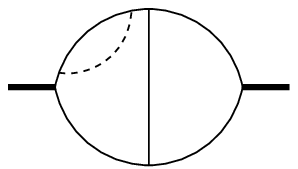}} &+& \raisebox{-5mm}{\epsffile{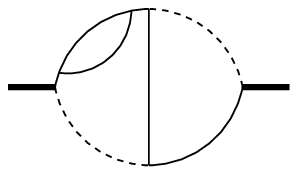}} &+2& \raisebox{-5mm}{\epsffile{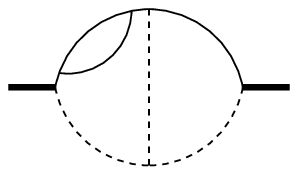}}
\vspace{2mm}
\\
+& \raisebox{-5mm}{\epsffile{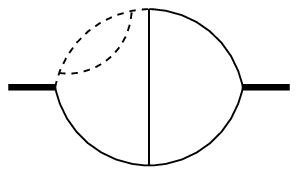}} &+2& \raisebox{-5mm}{\epsffile{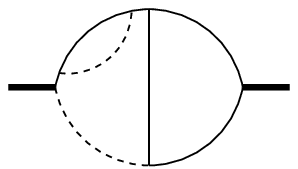}} &+2& \raisebox{-5mm}{\epsffile{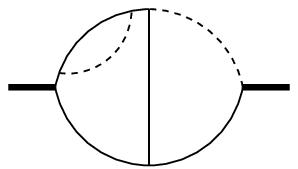}}
\vspace{2mm}
\\
+2& \raisebox{-5mm}{\epsffile{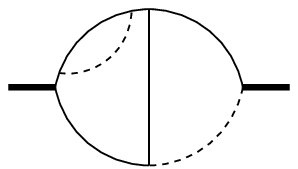}} &+2& \raisebox{-5mm}{\epsffile{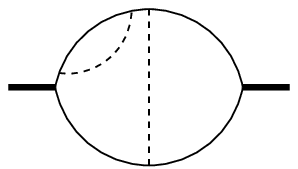}} &-2& \raisebox{-5mm}{\epsffile{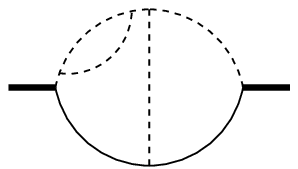}}
\vspace{2mm}
\\
-4& \raisebox{-5mm}{\epsffile{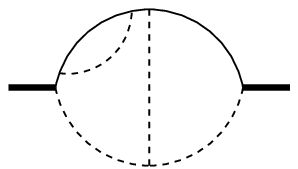}} &-& \raisebox{-5mm}{\epsffile{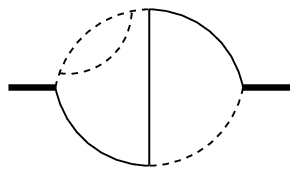}} &-2& \raisebox{-5mm}{\epsffile{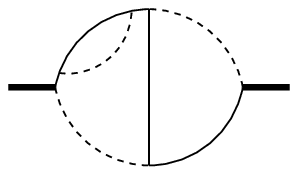}}
\end{array}
\end{eqnarray}
\begin{eqnarray}
\begin{array}{cccccc}
& \raisebox{-5mm}{\epsffile{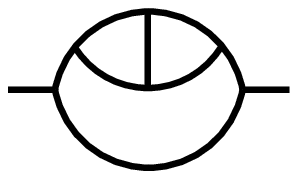}} &=& \raisebox{-5mm}{\epsffile{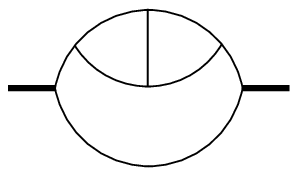}} &-4& \raisebox{-5mm}{\epsffile{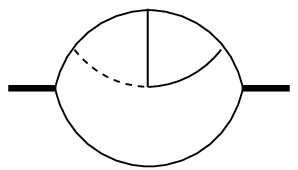}}
\vspace{2mm}
\\
-& \raisebox{-5mm}{\epsffile{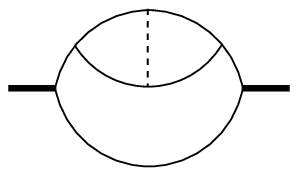}} &-& \raisebox{-5mm}{\epsffile{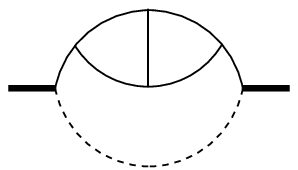}} &+2& \raisebox{-5mm}{\epsffile{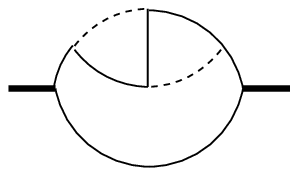}}
\vspace{2mm}
\\
+4& \raisebox{-5mm}{\epsffile{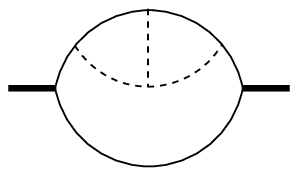}} &+4& \raisebox{-5mm}{\epsffile{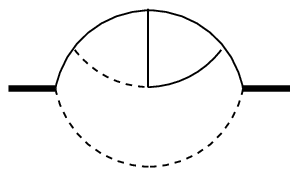}} &+& \raisebox{-5mm}{\epsffile{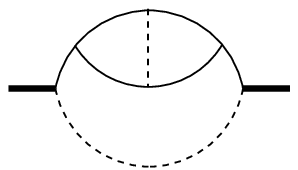}}
\vspace{2mm}
\\
-2& \raisebox{-5mm}{\epsffile{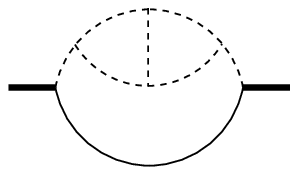}} &-2& \raisebox{-5mm}{\epsffile{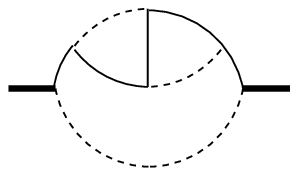}} &-4& \raisebox{-5mm}{\epsffile{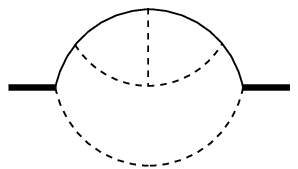}}
\end{array}
\end{eqnarray}
\begin{eqnarray}
\begin{array}{cccccc}
& \raisebox{-5mm}{\epsffile{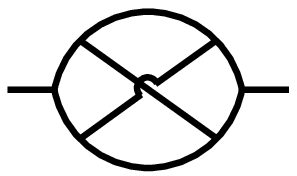}} &=& \raisebox{-5mm}{\epsffile{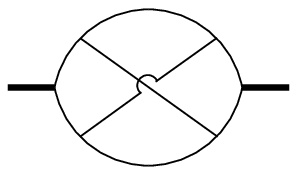}} &-4& \raisebox{-5mm}{\epsffile{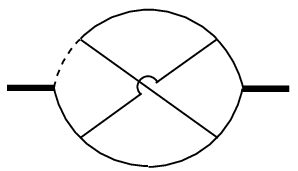}}
\vspace{2mm}
\\
-4& \raisebox{-5mm}{\epsffile{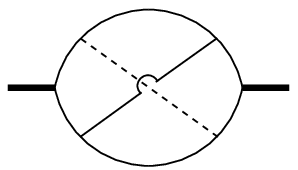}} &+8& \raisebox{-5mm}{\epsffile{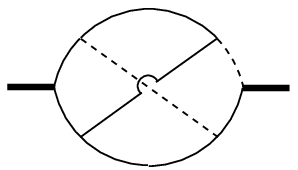}} &+4& \raisebox{-5mm}{\epsffile{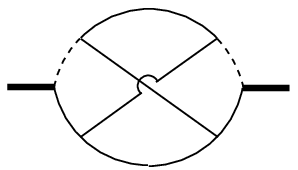}}
\vspace{2mm}
\\
+2& \raisebox{-5mm}{\epsffile{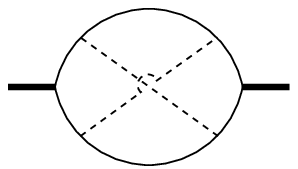}} &+8& \raisebox{-5mm}{\epsffile{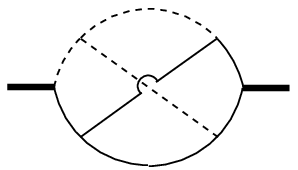}} &-16& \raisebox{-5mm}{\epsffile{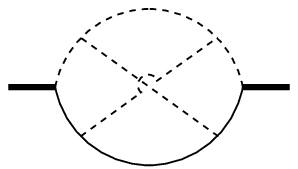}}
\vspace{2mm}
\\
-4& \raisebox{-5mm}{\epsffile{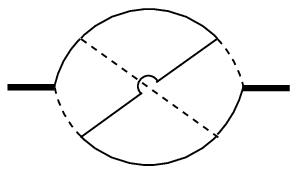}} & & & &
\end{array} 
\end{eqnarray}  
\begin{eqnarray}
\begin{array}{cccccc}
& \raisebox{-5mm}{\epsffile{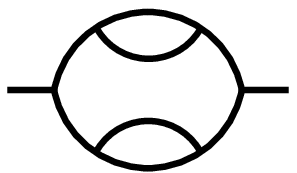}} &=& \raisebox{-5mm}{\epsffile{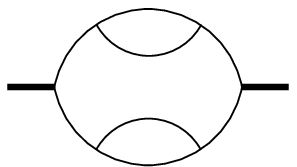}} &-4& \raisebox{-5mm}{\epsffile{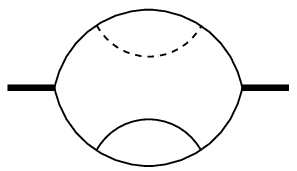}}
\vspace{2mm}
\\
+4& \raisebox{-5mm}{\epsffile{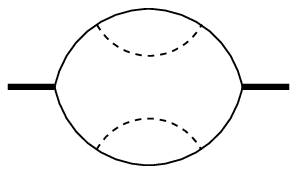}} &+2& \raisebox{-5mm}{\epsffile{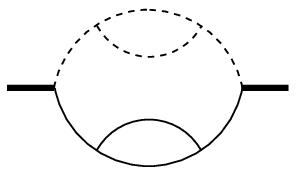}} &-4& \raisebox{-5mm}{\epsffile{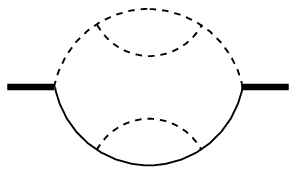}}
\end{array} 
\end{eqnarray}
\begin{eqnarray}
\begin{array}{cccccc}
 &\raisebox{-5mm}{\epsffile{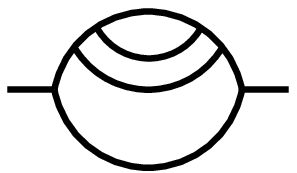}} &=& \raisebox{-5mm}{\epsffile{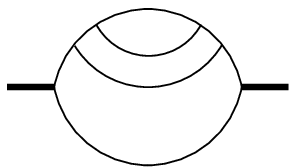}} &-& \raisebox{-5mm}{\epsffile{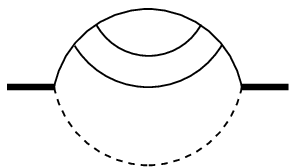}}
\vspace{2mm}
\\
-& \raisebox{-5mm}{\epsffile{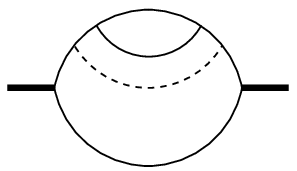}} &-2& \raisebox{-5mm}{\epsffile{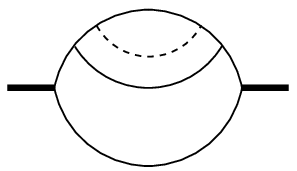}} &+2& \raisebox{-5mm}{\epsffile{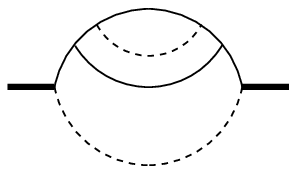}}
\vspace{2mm}
\\
+2& \raisebox{-5mm}{\epsffile{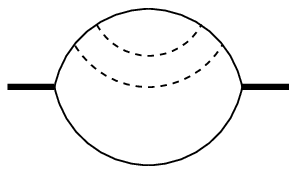}} &+& \raisebox{-5mm}{\epsffile{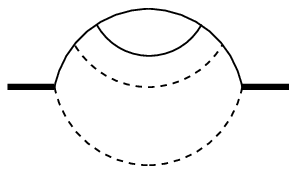}} &+& \raisebox{-5mm}{\epsffile{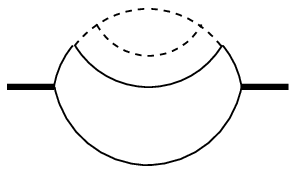}}
\vspace{2mm}
\\
-2& \raisebox{-5mm}{\epsffile{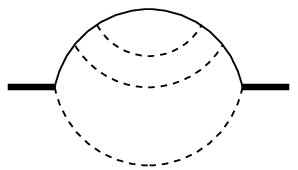}} &-& \raisebox{-5mm}{\epsffile{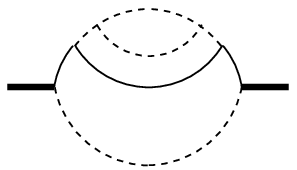}} &-& \raisebox{-5mm}{\epsffile{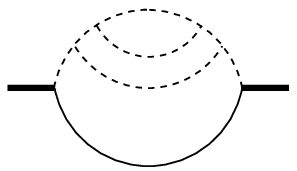}}
\end{array}
\end{eqnarray}
\begin{eqnarray}
\begin{array}{cccccc}
 & \raisebox{-5mm}{\epsffile{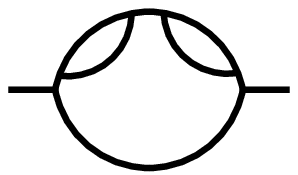}} &=& \raisebox{-5mm}{\epsffile{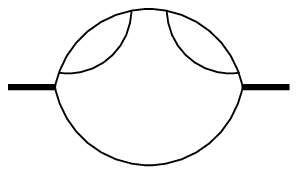}} &-4& \raisebox{-5mm}{\epsffile{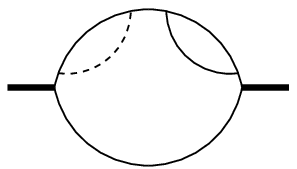}}
\vspace{2mm}
\\
-& \raisebox{-5mm}{\epsffile{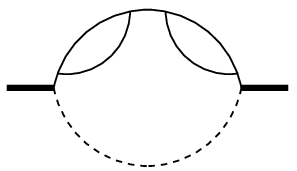}} &+4& \raisebox{-5mm}{\epsffile{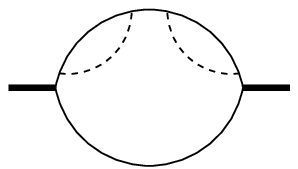}} &+4& \raisebox{-5mm}{\epsffile{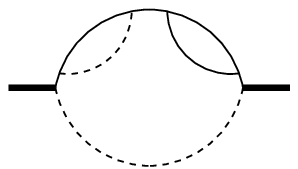}}
\vspace{2mm}
\\
-4& \raisebox{-5mm}{\epsffile{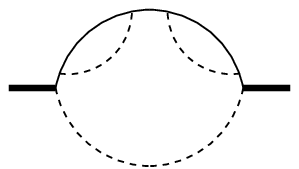}} &-& \raisebox{-5mm}{\epsffile{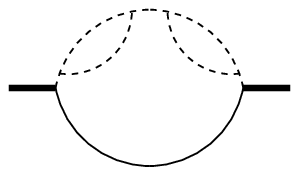}} & &
\end{array} 
\end{eqnarray}

\section{Diagrams For The Red Bonds}
\label{app:diagramsForReds}
This appendix gives details on the diagrammatic contributions to the renormalization of $w_r$ and $\tau$ in the limit $r \to \infty$. As an example we consider the one-loop diagrams A and B. In Sec.~\ref{redBonds} we argued, that only singly connected conducting propagators contribute to $Z_{w_\infty}$. Thus, A gives no such contribution at all. The contribution of B can be expressed as 
\begin{eqnarray}
\epsfxsize=2.3cm
\raisebox{-6mm}{\epsffile{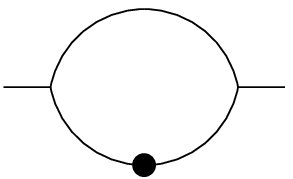}} \ ,
\end{eqnarray}
where the lines stand for conducting propagators evaluated at zero currents and the solid dot for an $\frac{1}{2} \phi^2$-insertion. The resulting contribution of $\mbox{A}-2\mbox{B}$ is
\begin{eqnarray}
\label{resA-2B}
\epsfxsize=2.3cm
-2 \ \raisebox{-6mm}{\epsffile{insertionOneLoop.eps}} \ .
\end{eqnarray}

Now we turn to $\tau$. $Z_\tau$ can be calculated by inserting $\frac{1}{2} \phi^2$ into conducting and insulating propagators. The contribution of both A and B reads
\begin{eqnarray}
\epsfxsize=2.3cm
2 \ \raisebox{-6mm}{\epsffile{insertionOneLoop.eps}} \ .
\end{eqnarray}
The resulting contribution of $\mbox{A}-2\mbox{B}$ is again the one stated in (\ref{resA-2B}).

We carry out the insertion procedure for both, $w_\infty$ and $\tau$, up to three-loop order. One obtains in both cases the same diagrams with the same fore-factors. The result is listed in Fig.~\ref{diagrammaticExpansionRedBonds}.
\section{Evaluation Of Diagrams For The Backbone}
\label{app:evaluation_r_-1}
In this appendix we give some details of the calculation of the backbone dimension. As described in Sec.~\ref{backbone}, we insert $\frac{1}{2} \phi^2$ into each conducting propagator. The diagram A for example has two conducting propagators. Its contribution to $Z_{w_{-1}}$ can be expressed as
\begin{eqnarray}
\epsfxsize=2.3cm
2 \ \raisebox{-6mm}{\epsffile{insertionOneLoop.eps}} \ .
\end{eqnarray}
As in App.~\ref{app:diagramsForReds}, the lines stand for conducting propagators evaluated at zero currents and the solid dot for a $\frac{1}{2} \phi^2$-insertion. The diagram B contributes via
\begin{eqnarray}
\epsfxsize=2.3cm
\raisebox{-6mm}{\epsffile{insertionOneLoop.eps}} \ ,
\end{eqnarray}
and hence the total contribution of $\mbox{A} - 2\mbox{B}$ vanishes.

The procedure is carried out up to three-loop order. It results in 
\begin{eqnarray}
\label{ideenlos}
\epsfxsize=2.3cm
\begin{array}{cccccc}
- & \raisebox{-5mm}{\epsffile{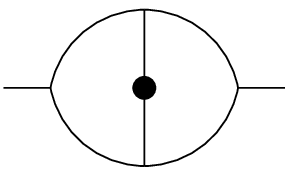}} & +2 & \raisebox{-5mm}{\epsffile{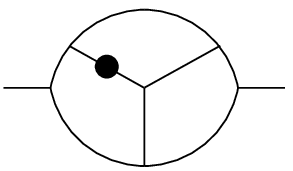}} & +2 &\raisebox{-5mm}{\epsffile{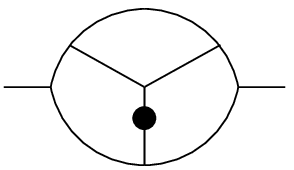}}
\vspace{2mm} 
\\
+4 & \raisebox{-5mm}{\epsffile{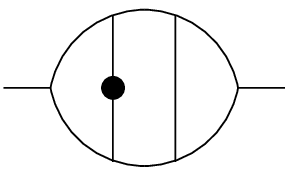}} & +2 & \raisebox{-5mm}{\epsffile{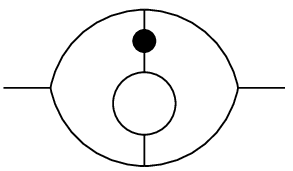}} & + & \raisebox{-5mm}{\epsffile{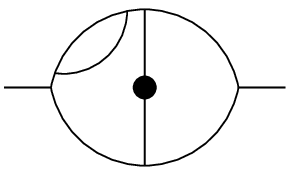}}
\vspace{2mm} 
\\
+8 & \raisebox{-5mm}{\epsffile{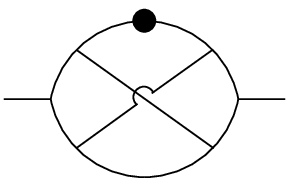}} & & & &
\end{array} \ .
\end{eqnarray}
These diagrams can be most conveniently evaluated by mapping them onto those calculated in\cite{alcantara_80}. The two-loop contribution for example can be re-expressed as
\begin{eqnarray}
\epsfxsize=2.3cm
-\frac{1}{2} \raisebox{-7mm}{\epsffile{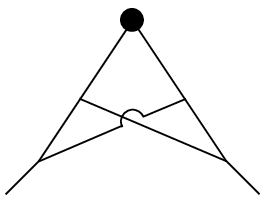}} = - \frac{1}{2} \left( \frac{-g}{2} \right)^{-1} \raisebox{-7mm}{\epsffile{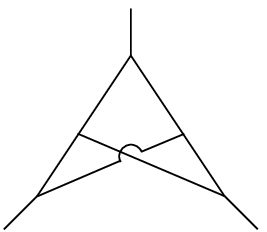}}   .
\end{eqnarray}
Note that we have explicitly extracted the combinatorial factor $\frac{1}{2}$ from the diagram. This is important at this stage, since different diagrams, each inheriting its combinatorial factor from its bold diagram, may be mapped onto the same three-leg diagram. The additional factor on the right hand side cancels the combinatorial factor $\frac{1}{2}$ of the three-leg diagram as well as a vertex $-g$. Similar identifications can be made for the three-loop diagrams appearing in Eq.~(\ref{ideenlos}). After all, the following diagrammatic contributions to the renormalization of $w_{-1}$ are obtained:
\begin{eqnarray}
\label{nochideeloser}
\epsfxsize=2.3cm
\begin{array}{cccccc}
- & \raisebox{-7mm}{\epsffile{mapping_1.eps}} & +\displaystyle{\frac{4}{3}}& \raisebox{-7mm}{\epsffile{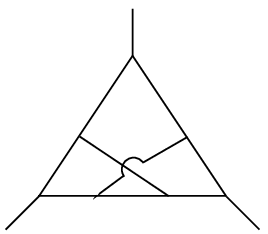}} & +2 & \raisebox{-7mm}{\epsffile{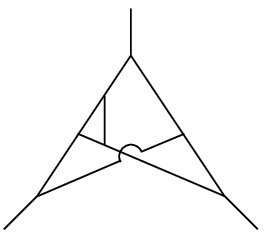}}
\vspace{2mm} 
\\
+\displaystyle{\frac{4}{3}}& \raisebox{-7mm}{\epsffile{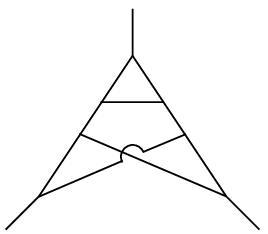}} & + &  \raisebox{-7mm}{\epsffile{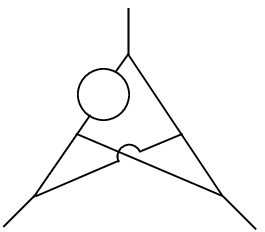}} & &
\end{array}  \ ,
\end{eqnarray}
where we have dropped an overall factor $- g^{-1}$.

The $\epsilon$-expansion results for the diagrams in Eq.~(\ref{nochideeloser}) can be gathered from\cite{alcantara_80}. However, we did not entirely rely on the results stated there. We also did the calculations on our own and found the same results leading to the renormalization factor given in Eq.~(\ref{zFactor}).
%

\newpage
\begin{figure}[h]
\end{figure}
\begin{figure}[h]
\epsfxsize=2.3cm
\begin{eqnarray*}
\begin{array}{cccccc}
-2 & \raisebox{-5mm}{\epsffile{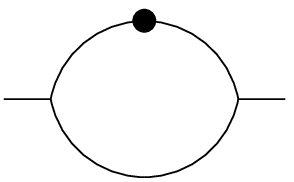}} & + & \raisebox{-5mm}{\epsffile{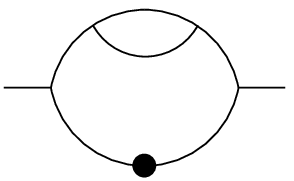}} & +2 & \raisebox{-5mm}{\epsffile{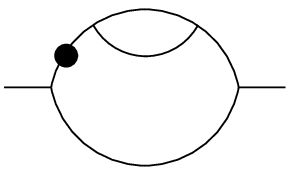}}
\vspace{2mm}
\\
+2 & \raisebox{-5mm}{\epsffile{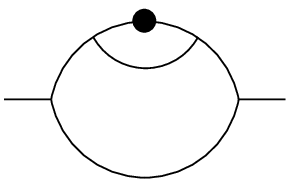}} & +8 & \raisebox{-5mm}{\epsffile{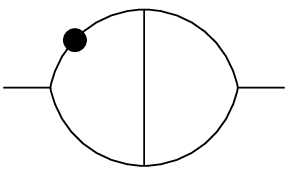}} & +2 & \raisebox{-5mm}{\epsffile{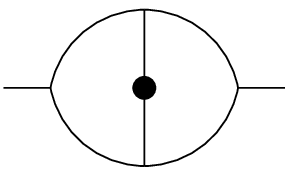}}
\vspace{2mm}
\\
-4 & \raisebox{-5mm}{\epsffile{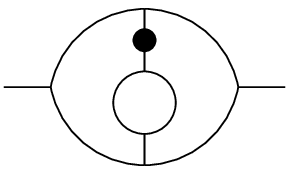}} & -8 & \raisebox{-5mm}{\epsffile{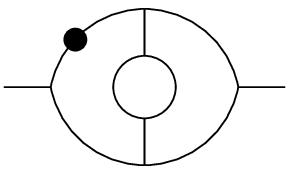}} & -4 & \raisebox{-5mm}{\epsffile{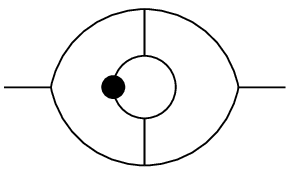}}
\vspace{2mm}
\\
-8 & \raisebox{-5mm}{\epsffile{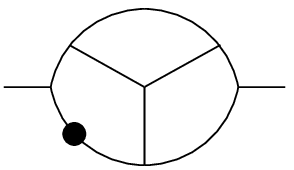}} & -8 & \raisebox{-5mm}{\epsffile{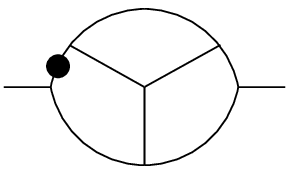}} & -8 & \raisebox{-5mm}{\epsffile{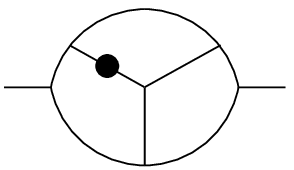}}
\vspace{2mm}
\\
-4 & \raisebox{-5mm}{\epsffile{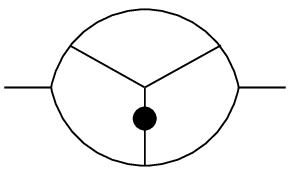}} & -4 & \raisebox{-5mm}{\epsffile{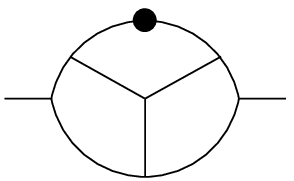}} & -16 & \raisebox{-5mm}{\epsffile{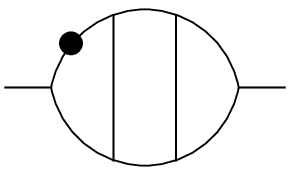}}
\vspace{2mm}
\\
-8 & \raisebox{-5mm}{\epsffile{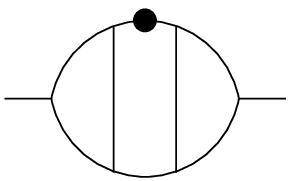}} & -8 & \raisebox{-5mm}{\epsffile{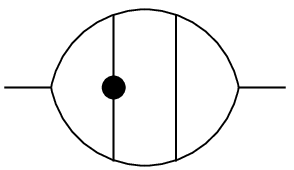}} & -4 & \raisebox{-5mm}{\epsffile{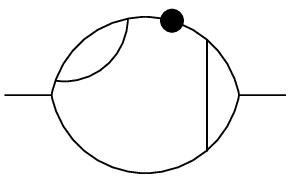}}
\vspace{2mm}
\\
-4 & \raisebox{-5mm}{\epsffile{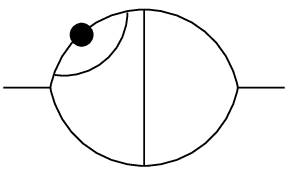}} & -2 & \raisebox{-5mm}{\epsffile{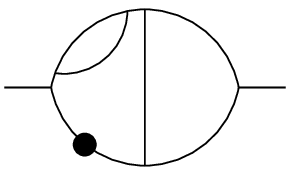}} & -2 & \raisebox{-5mm}{\epsffile{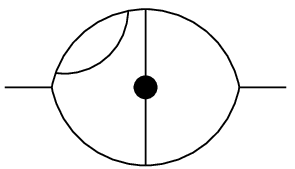}}
\vspace{2mm}
\\
-2 & \raisebox{-5mm}{\epsffile{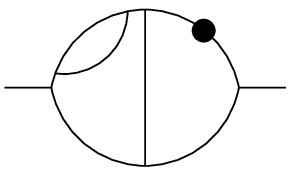}} & -2 & \raisebox{-5mm}{\epsffile{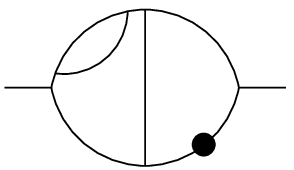}} & -4 & \raisebox{-5mm}{\epsffile{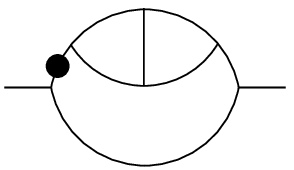}}
\vspace{2mm}
\\
-8 & \raisebox{-5mm}{\epsffile{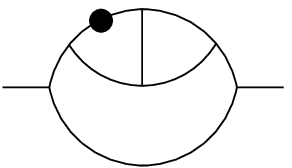}} & -2 & \raisebox{-5mm}{\epsffile{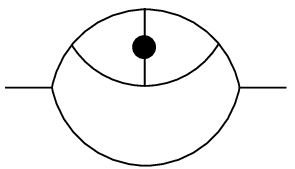}} & -2 & \raisebox{-5mm}{\epsffile{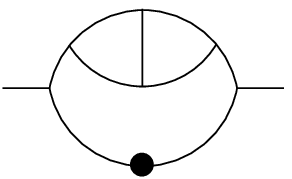}}
\vspace{2mm}
\\
-20 & \raisebox{-5mm}{\epsffile{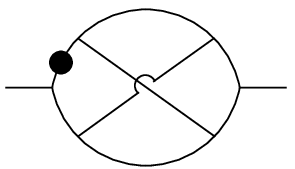}} & -20 & \raisebox{-5mm}{\epsffile{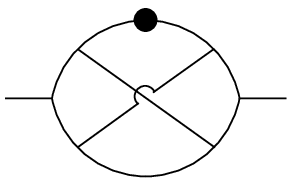}} & -4 & \raisebox{-5mm}{\epsffile{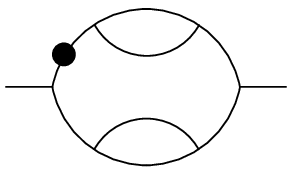}}
\vspace{2mm}
\\
-4 & \raisebox{-5mm}{\epsffile{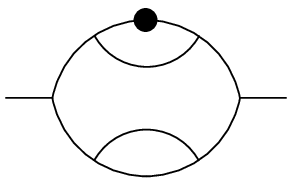}} & -2 & \raisebox{-5mm}{\epsffile{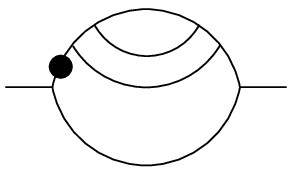}} & -2 & \raisebox{-5mm}{\epsffile{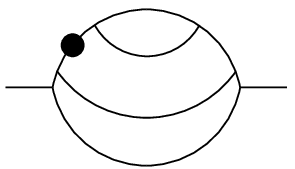}}
\vspace{2mm}
\\
-2 & \raisebox{-5mm}{\epsffile{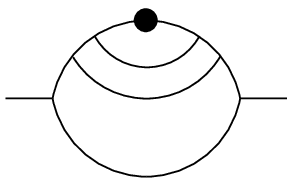}} & - & \raisebox{-5mm}{\epsffile{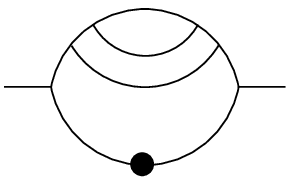}} & - & \raisebox{-5mm}{\epsffile{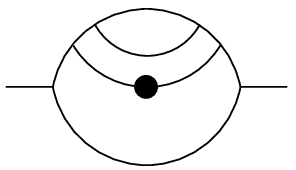}}
\vspace{2mm}
\\
-3 & \raisebox{-5mm}{\epsffile{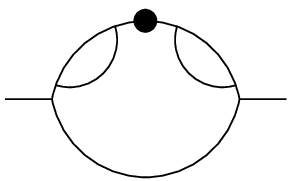}} & -4 & \raisebox{-5mm}{\epsffile{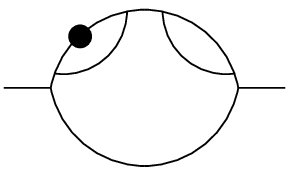}} & - & \raisebox{-5mm}{\epsffile{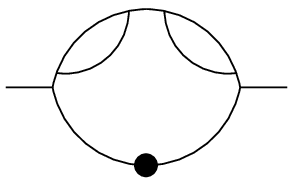}}
\vspace{2mm}
\end{array}
\end{eqnarray*}
\caption[]{\label{diagrammaticExpansionRedBonds}}
\end{figure}
\begin{figure}[h]
\epsfxsize=8.4cm
\centerline{\epsffile{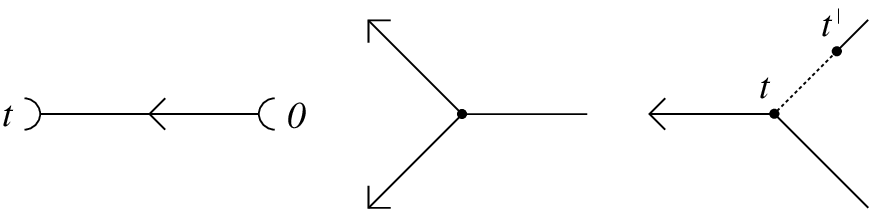}}
\caption[]{\label{newElements}}
\end{figure}
\begin{figure}[h]
\epsfxsize=2.4cm
\centerline{\epsffile{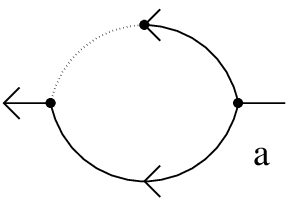}}
\caption[]{\label{dynamicOneLoop}}
\end{figure}
\begin{figure}[h]
\epsfxsize=2.3cm
\begin{eqnarray*}
\begin{array}{cccccc}
&\raisebox{-6mm}{\epsffile{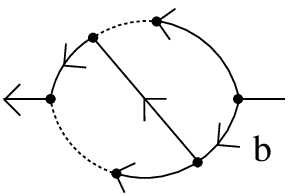}} &+& \raisebox{-6mm}{\epsffile{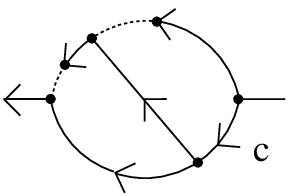}} &+& \raisebox{-6mm}{\epsffile{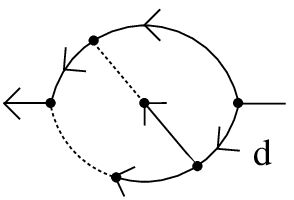}}
\vspace{2mm}
\\
+ & \raisebox{-6mm}{\epsffile{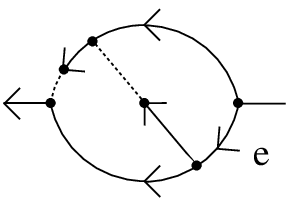}} & & & &
\end{array}
\end{eqnarray*}
\caption[]{\label{dynamicTwoLoop1}}
\end{figure}
\begin{figure}[h]
\epsfxsize=2.3cm
\begin{eqnarray*}
\begin{array}{cccccc}
&\raisebox{-6mm}{\epsffile{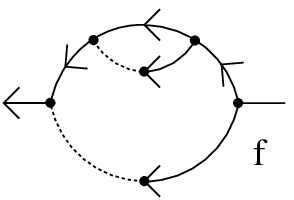}} &+& \raisebox{-6mm}{\epsffile{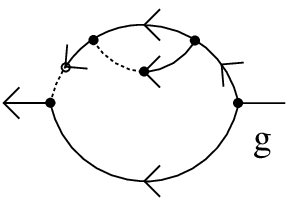}} & &
\end{array}
\end{eqnarray*}
\caption[]{\label{dynamicTwoLoop2}}
\end{figure}
\begin{figure}[h]
\epsfxsize=8.4cm
\centerline{\epsffile{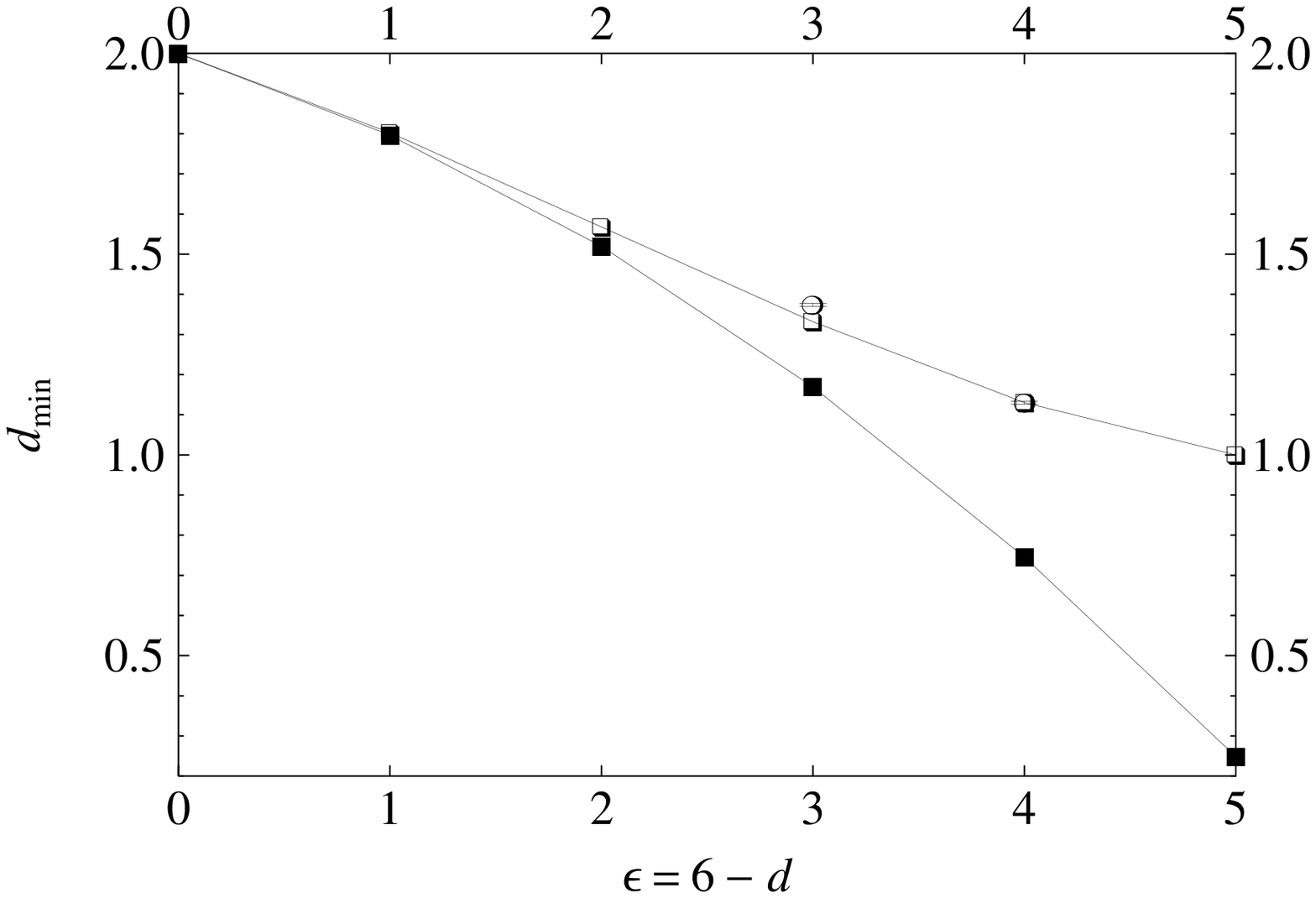}}
\centerline{\epsffile{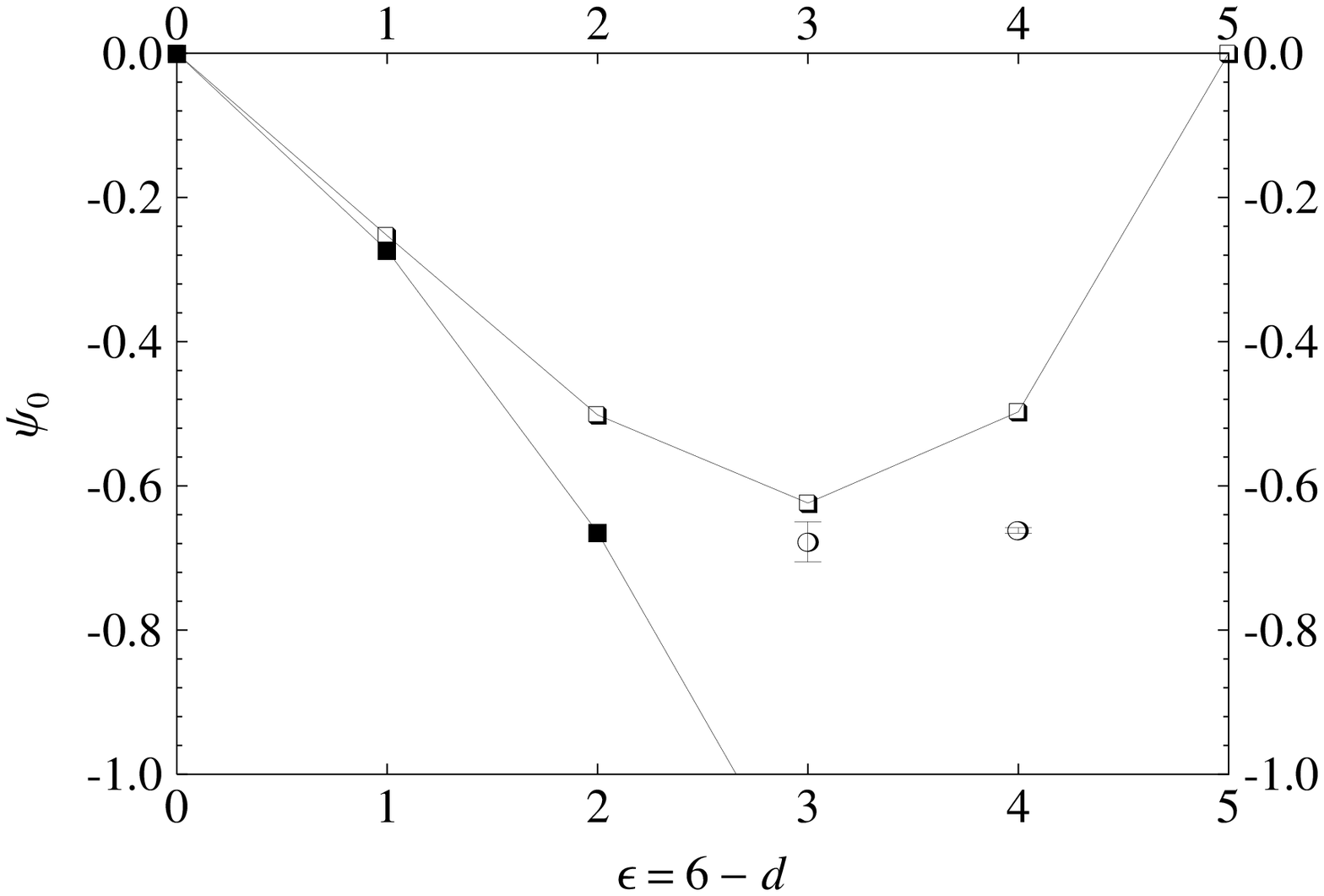}}
\caption[]{\label{dataDmin}}
\end{figure}
\begin{figure}[h]
\epsfxsize=8.4cm
\centerline{\epsffile{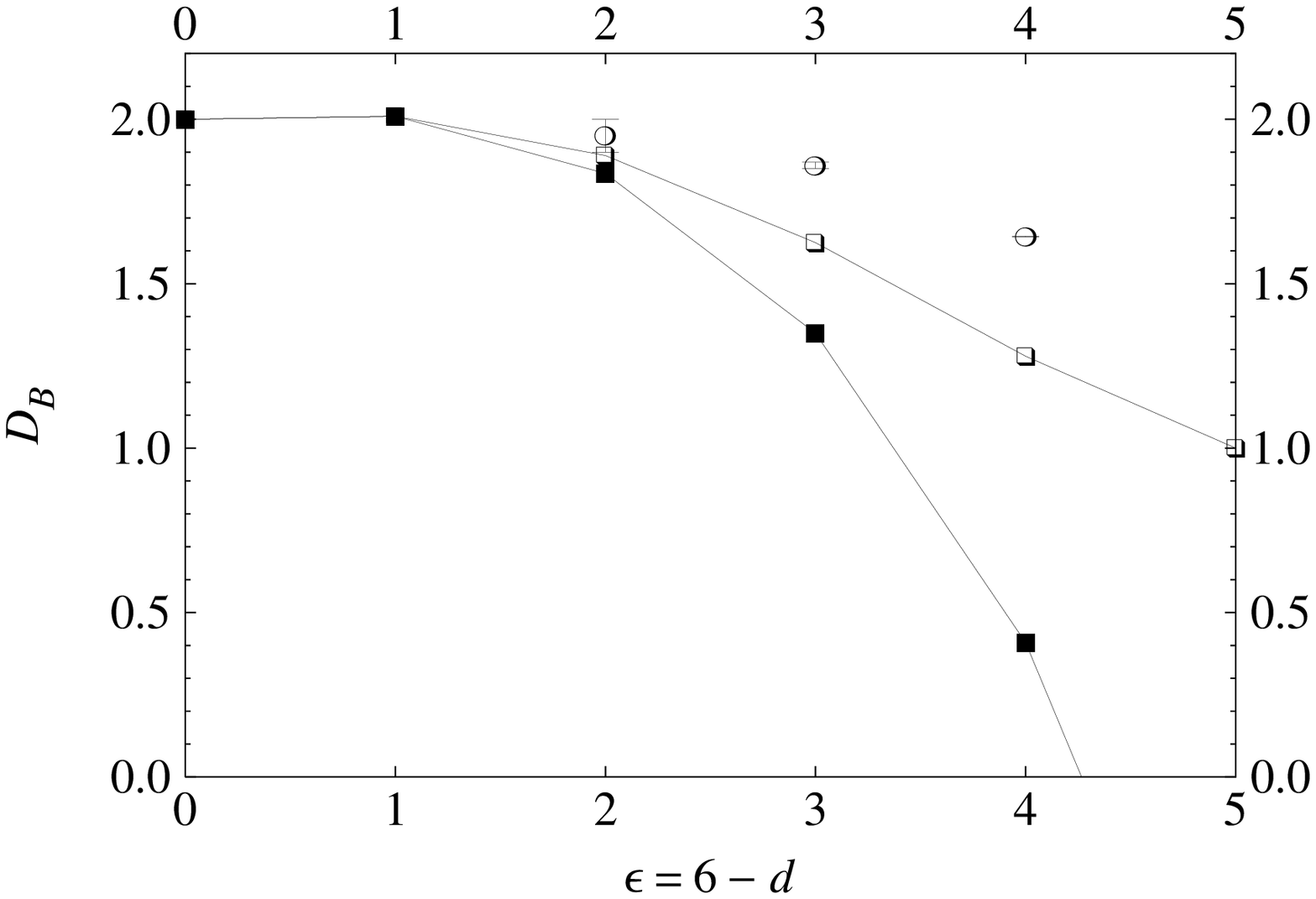}}
\centerline{\epsffile{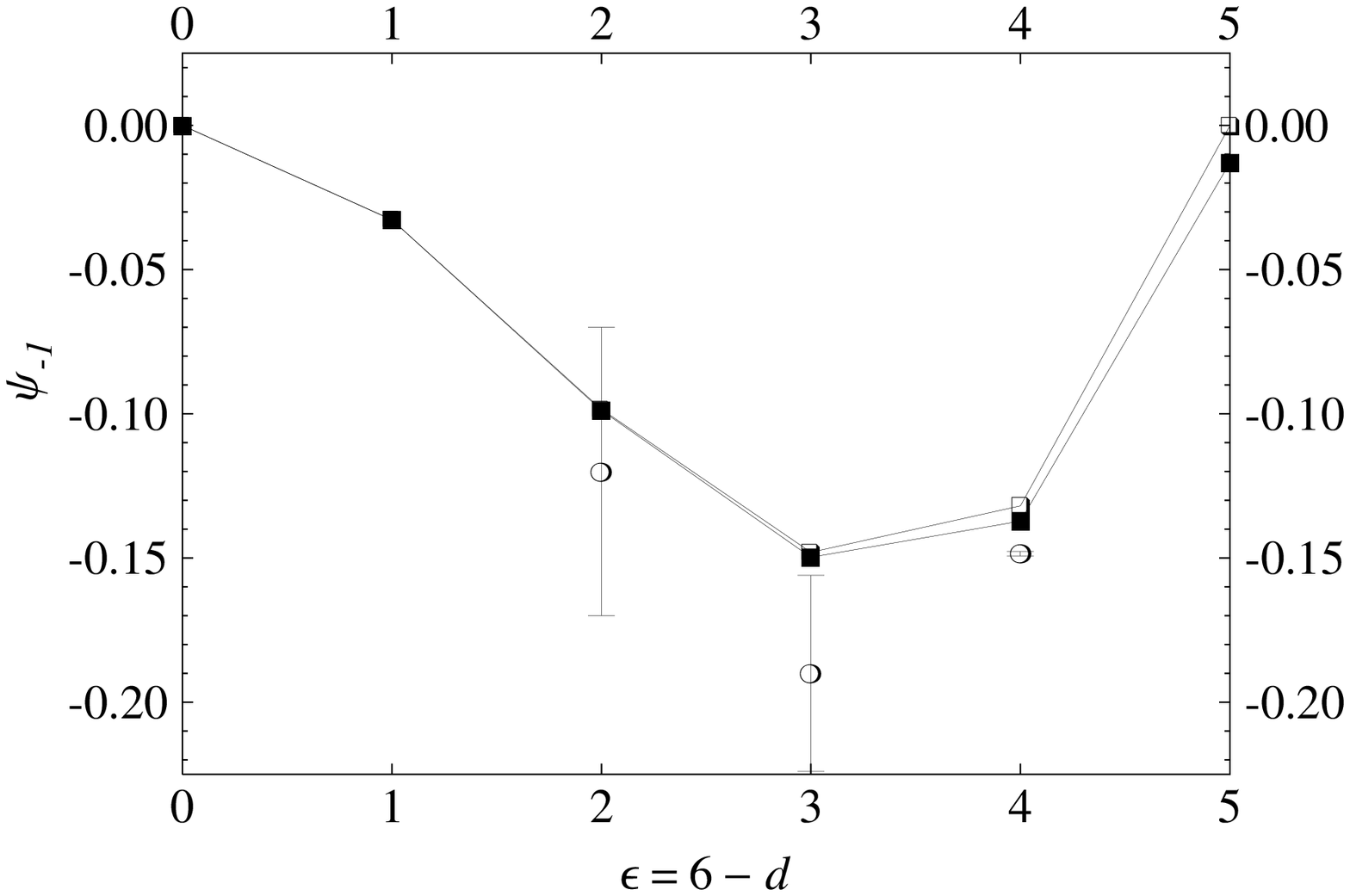}}
\caption[]{\label{dataDb}}
\end{figure}

\noindent
FIG.~\ref{diagrammaticExpansionRedBonds}\newline
Diagrammatic expansion in the limit $r \to \infty$. The listed diagrams including their fore-factors can be obtained from the conducting diagrams shown in App.~\ref{app:decomposition} in two different ways: firstly, by inserting $\frac{1}{2} \varphi^2$ into all singly connected conducting propagators and secondly, by inserting $\frac{1}{2} \varphi^2$ into every conducting and insulating propagator. As a consequence, the renormalization factors $Z_{w_\infty}$ and $Z_\tau$ are identical. The lines stand for conducting propagators evaluated at zero currents, the solid dots for $\frac{1}{2} \varphi^2$-insertions.
\newline\vspace{0.4cm}\\
FIG.~\ref{newElements}\newline
The propagator $G \left( {\rm{\bf p}},t \right)$ as well as the vertices $\gamma g$ and $-\gamma g \theta \left( t - t^\prime \right)$ (from left to right).
\newline\vspace{0.4cm}\\
FIG.~\ref{dynamicOneLoop}\newline
In the limit $r \to 0^+$ we map the bold one-loop diagram (see 
App.~\ref{app:decomposition}) onto the dynamic one shown here. The meaning of 
the graphic elements may be inferred from Fig.~\ref{newElements}.
\newline\vspace{0.4cm}\\
FIG.~\ref{dynamicTwoLoop1}\newline
Dynamic diagrams obtained in the limit $r \to 0^+$.
\newline\vspace{0.4cm}\\
FIG.~\ref{dynamicTwoLoop2}\newline
Dynamic diagrams obtained in the limit $r \to 0^+$.
\newline\vspace{0.4cm}\\
FIG.~\ref{dataDmin}\newline
Dependence of the exponents $d_{\mbox{{\scriptsize min}}}$ and $\psi_{0} = 
d_{\mbox{{\scriptsize min}}} - \gamma /\nu$ on dimensionality. The 
$\epsilon$-expansion (full squares) and the rational approximation (open 
squares) are compared to numerical simulations (circles). For 
$d_{\mbox{{\scriptsize min}}}$ we take Monte Carlo results by Grassberger. At $d=2$ we insert the exact values\cite{nijs_79,nienhuis_82} $\nu =4/3$ and $\gamma =43/18$. At $d=3$ we use Monte Carlo results by Ziff and Stell\cite{ziff_stell}:  $\nu = 0.875 \pm 0.008$, $\gamma =1.795 \pm 0.005$.
\newline\vspace{0.4cm}\\
FIG.~\ref{dataDb}\newline
Dependence of the exponents $D_B$ and $\psi_{-1} = D_B - \gamma /\nu$ on 
dimensionality. The $\epsilon$-expansion (full squares) and the rational 
approximation (open squares) are compared to numerical results (circles) by 
Grassberger ($d=2$) and Moukarzel ($d=3,4$). They determined $D_B$ by 
simulations. At $d=2$ and $d=3$ we use the same values for $\gamma$ and $\nu$ as in Fig.~\ref{dataDmin}. At $d=4$ we take $\nu^{-1} = 1.44 \pm 
0.05$\cite{moukarzel_98} and $\gamma =1.44$\cite{stauffer_aharony_92}.
%
%
\newpage
\newpage
\newpage
\end{document}